\documentclass{aa}  
\usepackage{graphicx}
\usepackage{natbib}
\usepackage{txfonts}
\bibpunct{(}{)}{;}{a}{}{,}
\begin{document}
   \title{Analysis of extremely low signal-to-noise ratio data 
   from INTEGRAL/PICsIT}


   \author{P. Lubi\'nski \inst{1,2}}

   \offprints{P. Lubi\'nski}

   \institute{Centrum Astronomiczne im. M. Kopernika, 
              Bartycka 18, PL-00-716 Warszawa, Poland
         \and
             ISDC Data Centre for Astrophysics, Chemin d'Ecogia 16,
	     CH-1290 Versoix, Switzerland\\
         \email{Piotr.Lubinski@unige.ch}
             }

   \date{Received ; accepted }

 
  \abstract
   {The PICsIT detector onboard the INTEGRAL satellite was designed to provide 
   information about emission in the soft $\gamma$-ray band for many bright 
   sources. Due to strong and variable instrumental background, only 4 
   objects have been detected so far using standard software.}  
   {The moderate sensitivity of PICsIT can be compensated for in the case of 
   many objects by adopting a long exposure time, thanks to INTEGRAL's large 
   field of view and an observing strategy focused on the Galactic plane. With 
   angular resolution far higher than that of all other instruments operating 
   in a similar energy band, PICsIT is suitable for fields too crowded or too 
   significantly affected by Galactic diffuse emission. Therefore, it is 
   desirable to improve the spectral extraction software to both obtain more
   reliable results and enlarge the number of objects that can be studied.}
   {The new PICsIT spectral extraction method is based on three elements: 
   careful modelling of the background, an energy-dependent pixel-illumination 
   function, and the computation of the probability density  of the 
   source count rate. Background maps of the detector plane are prepared for 
   short periods and relatively narrow energy bands to insure that the 
   background dependence on time and energy is modelled well. The most 
   important element of the new spectral extraction method is the proper
   treatment of the Poisson-distributed data, developed within a Bayesian 
   framework.}
   {The new method was tested extensively on both a large true data set 
   and simulated data. Results assumed in simulations were reproduced perfectly,
   without any bias and with high precision. Count rates measured for Crab
   were far more stable than those obtained with the standard software. For 
   weaker sources, the new method produced spectra of far higher quality and 
   allows us to detect at least  8  additional objects. Comparison with other 
   INTEGRAL instruments demonstrated that PICsIT is well calibrated and provides  
   valuable information about the continuum emission in the 250 keV -- 1 MeV 
   band, detectable currently only by INTEGRAL.}
   {}

	   
   \titlerunning{Analysis of INTEGRAL/PICsIT data}

   \maketitle
%

\section{Introduction}

The low-energy $\gamma$-ray band (100 keV -- 10 MeV) is extremely difficult to 
observe, mainly because of the intense nuclear-decay emission induced by cosmic 
rays inside the detector and surrounding materials. As the flux of the observed 
objects decline significantly with increasing energy from X-rays to 
$\gamma$-rays, the strong increase in internal background of the detector  
makes the ratio of observed signal to background very small. This ratio 
usually does not exceed 0.01 even for sources as bright as the 
Crab nebula, despite the passive and active shielding applied. Moreover,  the 
fact that  soft $\gamma$-rays cannot be focused as well as soft X-rays reduces 
the angular resolution in this energy range. Therefore, observations of crowded 
Galactic fields with $\gamma$-ray telescopes are often affected by contamination 
from nearby sources or from Galactic diffuse emission.

There have been many $\gamma$-ray observatories, but due to their generally 
limited life times and sensitivities only a few have provided a significant 
amount of data. More detailed studies of the soft $\gamma$-ray spectra became 
possible in the last decade of the 20th century, thanks to the SIGMA detector 
\citep{Paul91} onboard the GRANAT satellite, and the OSSE and BATSE detectors 
\citep{Johnson93} onboard the CGRO satellite. SIGMA was a coded-mask instrument 
with a good angular resolution of 13$\arcmin$, and its nominal energy range was 
35--1300 keV. The OSSE telescope has been the most sensitive soft $\gamma$-ray 
instrument sent into orbit until now, due to its large size. The high sensitivity 
of OSSE was in practice slightly reduced by limitations related to its low orbit. 
In addition, the large field of view of OSSE \rm made it \rm difficult to resolve 
closeby objects at the Galactic centre and to distinguish point-source emission 
from Galactic diffuse emission. BATSE was an all-sky monitor, used mainly for 
detecting GRBs and monitoring the activity of brighter $\gamma$-ray sources. The 
nominal energy range of the PDS detector \citep{Frontera97} onboard the BeppoSAX 
satellite was 15--300 keV but its sensitivity above 150 keV was insufficient to 
provide more precise data. 

There are four missions presently in operation with soft $\gamma$-ray 
instruments, namely RXTE (HEXTE), INTEGRAL, Swift (BAT), and Suzaku (HXD/GSO). 
The HEXTE \citep{Rothschild98} and BAT \citep{Gehrels04} energy ranges do not 
exceed $\approx$150 keV, and these instruments have therefore limited 
capabilities of studying the high energy emission. GSO onboard Suzaku was 
declared to be the detector with the lowest internal background in the 40--600 
keV band, due to a more advanced active-shielding technique \citep{Takahashi07}. 
This low background should correspond to high sensitivity, but 3 years after 
Suzaku's launch almost no results above 200 keV have been published, apart from
Cyg X-1 observed up to 400 keV \citep{Makishima08}. Therefore, taking into 
account the limited angular resolution of GSO above 100 keV ($\approx$ 
4.5$\degr$), current interest in the soft $\gamma$-ray research is focused on 
the data acquired by INTEGRAL detectors.

The INTEGRAL satellite \citep{Winkler03} has onboard three coded-mask 
$\gamma$-ray detectors. ISGRI \citep{Lebrun03} is the upper detector of 
INTEGRAL's imager IBIS \citep{Ubertini03}, which operates nominally in the 
13--1000 keV range. Owing to its relatively high sensitivity and angular 
resolution of 12$\arcmin$, ISGRI remains the INTEGRAL instrument of choice for 
studies of point sources, particularly Galactic ones. The lower layer of the IBIS 
imager, the PICsIT detector \citep{Labanti03}, is described in the next section. 
The spectrometer SPI \citep{Vedrenne03} is a high spectral resolution 
$\gamma$-ray telescope operating in the energy range 20 keV -- 8 MeV. Despite its 
lower sensitivity and limited angular resolution ($\approx$ 2.5$\degr$), SPI 
spectra extracted for brighter sources complement the ISGRI spectra well, 
especially above 150 keV where the ISGRI sensitivity declines rapidly. Due to its
reliable, ground-based efficiency calibration, SPI serves also as a reference 
instrument for the INTEGRAL cross-calibration tests. Observations of crowded 
fields, however, are difficult, as for OSSE, and special care is required in 
analysing such data \citep{Roques05}. 

Notwithstanding their unique capabilities, SPI and ISGRI provide rather limited
information about the continuum emission from point sources above $\approx$200 
keV. In the case of SPI, besides its limited angular resolution, there are some 
difficulties in modelling the instrumental background at higher energy, where
the number of photons is low. However, by applying a more sophisticated analysis 
to SPI data, it was possible to detect 20 objects above 200 keV 
\citep{Bouchet08}. The sensitivity of ISGRI declines dramatically above 150 keV, 
and the efficiency calibration remains uncertain in that range 
\citep{Jourdain08}. 

Initially, the INTEGRAL imager sensitivity was planned to exceed that of OSSE by 
about an order of magnitude \citep{Winkler94}. Unfortunately, due to a reduction 
in mission funds, this plan became impossible, when the foreseen three layers of 
the high energy detector were reduced to two layers of smaller volume and 
a limited telemetry was ascribed to them. The effect is that the PICsIT detector,
being still the largest soft $\gamma$-ray detector on orbit, has the thickness of 
only 3 cm compared to almost 18 cm of OSSE. The sensitivity of PICsIT is reduced 
further because the pixellated structure and data-acquisition logic produce an
additional loss in the fraction of the Compton-scattered photons. On the other 
hand, due to the elongated orbit, both the Earth occultations and passages 
through the radiation belts affect only a small part of the INTEGRAL revolution 
period. This and the observing strategy focused on the Galactic centre and 
Galactic plane ensures that many objects are observed by PICsIT with a long 
exposure time, compensating to some extent the moderate sensitivity of the 
instrument.

Shortly after launch, it appeared that the  PICsIT background level was about 
two times lower than the pre-launch conservative estimate \citep{DiCocco03}. 
Only the energy band below 300 keV showed a higher background, with an excess 
originating in track events induced by high-energy cosmic rays \citep{Segreto03}. 
Despite the lower background, the sensitivity achieved after the first year of 
detector operation was low, a factor between 5 and 10 lower than the statistical 
limit. The situation was improved considerably when the long-exposure background 
maps were included in the OSA 4.0 software release \citep{Foschini04}, allowing 
a sensitivity comparable to that of SPI to be achieved.

Nonetheless, almost six years after the INTEGRAL launch, there remain only three 
objects for which results with PICsIT have been reported: Crab \citep{DiCocco03}, 
Cyg X-1 \citep{Cadolle06}, and XTE J1550-564 during its 2003 outburst 
\citep{Foschini05}. The PICsIT spectra presented did not provide significantly
more information than the SPI spectra. More spectacular PICsIT results were 
instead obtained for the observations of gamma-ray bursts (GRBs) 
\citep{Malaguti03b}, in particular those  appearing outside the INTEGRAL field of 
view \citep{Marcinkowski06}, because these events are detectable at high energy 
with the use of IBIS Compton-mode data. 

Since the possibilities of improving the spectral extraction performance for
PICsIT using the standard OSA software were limited, a completely novel approach 
was developed for testing purposes. Preliminary results demonstrated that this 
new approach produced very good results, by for example allowing us
to detect several objects not detected by the standard procedure. This paper 
presents all the elements of the new method, which is now fully developed. Since 
we decided after discussions with the IBIS Team that the new method should not 
be implemented in the  standard software for INTEGRAL data analysis, we provided
a detailed presentation here to allow a potential user to reproduce the results. 
In Sects. \ref{instrument}--\ref{ppdf}, basic ingredients of the spectral 
extraction technique are described. Section \ref{limits} presents the results of 
testing the PICsIT detection limits. In Sect. \ref{spectra}, our main results, 
PICsIT spectra, are widely presented and discussed. After a summary given in 
Sect. \ref{summary}, an extended Appendix A presents the results of tests 
completed for different methods that can be used to extract the source count 
rates from $\gamma$-ray instruments.  


\section{Instrument model}
\label{instrument}

A detailed description of the PICsIT detector and IBIS in general can be found in 
the IBIS Observer's Manual \citep{Kuulkers06} and in the references listed there. 
We present here only basic information related to the pixel-illumination 
modelling.

PICsIT is the PIxellated Caesium Iodide Telescope, which operates in the nominal 
energy range of 175 keV -- 10 MeV and has an angular resolution of 12$\arcmin$. 
The fully coded field of view is a rectangle of 9$\degr \times$ 9$\degr$ and the 
50\% partially coded field of view has a size of 19$\degr \times$ 19$\degr$.  
The detector consists of 4096 CsI crystals, organized in 8 Modular Detection 
Units. Each pixel has the surface area of 8.55x8.55 mm$^{2}$ and a 30 mm 
thickness; pixels in a module are separated by 0.55 mm. The total sensitive area 
is 2994 cm$^{2}$. The IBIS mask is located 3283 mm above the PICsIT surface. The 
mask consists of 9025 open or closed cells, each of surface area 11.2x11.2 
mm$^{2}$, the total size of the mask being 1064x1064 mm$^{2}$. Closed cells 
consist of 16 mm thickness of tungsten. The ISGRI detector made of 16384 CdTe 
pixels (4x4x2 mm$^{3}$) is mounted 9 cm above the PICsIT surface. The IBIS 
collimator operates as a passive lateral shield and consists of the main IBIS 
tube, a hopper giving an additional limit to solid angle observable by detectors, 
and a side-mask shielding between the mask and tube walls. Both ISGRI and PICsIT 
detectors are surrounded with lateral and bottom BGO crystals that form an 
anti-coincidence shield, reducing the instrumental background by about a factor 
of 2.

PICsIT operates in three modes: photon-by-photon, spectral-imaging, and 
spectral-timing. Due to the limited INTEGRAL telemetry, photon-by-photon mode 
is in practice used only for tests. Therefore, almost all PICsIT data are stored 
as spectral-imaging histograms and spectral-timing tables. The spectral-imaging 
is the basic mode since it provides full position information (count rate 
for each pixel) and 256 channels of energy information. Histograms correspond to 
a single science window (a single spacecraft pointing), thus the time 
resolution of the spectral-imaging data is usually in the range of 30--120 
minutes. Spectral-timing data have a time resolution fixed between 1 and 500 ms 
but the count rates are integrated over all detector pixels and stored in 
only 8 energy channels. Therefore, this mode can be used only for extremely 
bright sources, typically GRBs, that have the signal strength that exceeds 
the background emission in short time intervals. 

The spectral-imaging histograms are made for two types of events: single 
and multiple. A single event detection is one in which there is only one photon 
in a single PICsIT pixel. Multiple events are defined to be detections of 2 or 3 
photons in a single submodule (half of the module, with pixels coupled 
electronically); the event position is then given by the position of the pixel 
with the highest energy deposit and the energy of the event equals the summed 
energy of all photons registered within the event time. The fraction of multiple 
events varies from $\approx$ 10\% in the 300--400 keV band to $\approx$ 50\% 
in the 1500--2200 keV band.  The logic of selecting single and multiple events 
based on one submodule detection introduces some non-uniformity because the 
multiple events in the border pixels are often registered as single events. This 
paper describes the spectral extraction applied to single events 
only, although the same procedure can be applied to multiple events.

The standard analysis software for all INTEGRAL data is the Off-line Science 
Analysis (OSA) package, distributed by the INTEGRAL Science Data Centre 
\citep{Courvoisier03}. The OSA software package presently allows for a 
limited PICsIT data analysis: image analysis for spectral-imaging data, and 
detector light curve and spectrum analysis for spectral-timing data. Taking into 
account that each source detected by PICsIT should be observable at lower 
energies by ISGRI, PICsIT alone is not expected to detect new sources. Therefore, 
the spectral extraction based on the Pixel Illumination Function (PIF) computed 
for the catalog source position appears to be more appropriate for quantitative 
PICsIT data analysis than the image deconvolution. The PIF-based method directly 
describes the  photon absorption that on the detector surface creates a shadow of 
the mask illuminated by the source observed at a given angle. In contrast, image 
deconvolution is an indirect method, which usually applies some type of inversion 
or cross-correlation that can lead to decreased efficiency. 

\begin{figure}[!h]
\centering
\includegraphics[width=9cm]{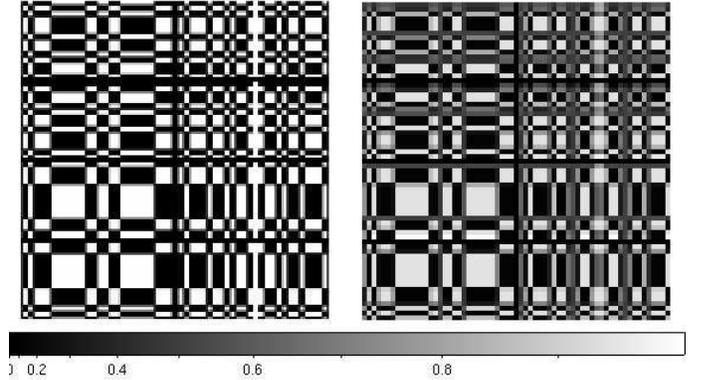}
\caption{ISGRI (left, 50 keV) and PICsIT (right, 450 keV) PIF models for science 
window 007900330010 calculated for Cyg X-1 seen at 3.9$\degr$ off-axis angle and
224$\degr$ azimuthal angle.  Azimuthal angle is measured anticlockwise from 
the right. The  PICsIT PIF is slightly shifted 
towards the upper right because of larger distance between the mask and detector. 
Fully open and fully closed elements transparency is respectively 1 and 0.003 for 
ISGRI and 0.89 and 0.11 for PICsIT.}
\label{pifs}
\end{figure}

There has been a spectral extraction component in the OSA software since version 
5.0 but the PIF model included is not yet reliable \citep{Foschini07c}. In 
consequence, all PICsIT spectra made with OSA are based on count rates determined 
from the mosaic images. For the purpose of the research presented here, 
a completely new software was developed for PIF modelling. The code was an 
adaptation of the code used to model the absorption in the Nomex structure 
supporting the IBIS mask \citep{Lubinski07}. The absorption of photons in the 
mask and the ISGRI detector layer was computed to be the mean fraction of source 
photons arriving at each pixel at given polar and azimuthal angles. Attenuation 
coefficients for tungsten (mask) and CdTe (ISGRI) are taken from the tables of 
\citet{Hubbell96}. The PIF map is calculated for 22 energy bins, with an 
effective energy for each bin that is calculated by assuming that the spectral 
index of the source emission is 2.1. The probability that the photon arrives at 
each pixel of the detector is computed over a grid where photons are separated by 
0.125 mm. The photon path length inside a given mask or ISGRI pixel was 
determined by numerical integration and the vertical step was also set to the 
value 0.125 mm, which was far smaller than the pixel size and thickness. Figure 
\ref{pifs} presents an example of ISGRI and PICsIT PIF models calculated for the 
same science window and the same source. The Nomex structure supporting the IBIS 
mask absorbs between 10\% and 25\% of the source photons at energies above 200 
keV, depending on the source off-axis angle. To correct the source count rates for 
this effect, a set of high-energy off-axis correction maps was prepared for 
PICsIT, in the same manner as those provided for ISGRI in the OSA package.

\begin{figure}[!h]
\centering
\includegraphics[width=5cm]{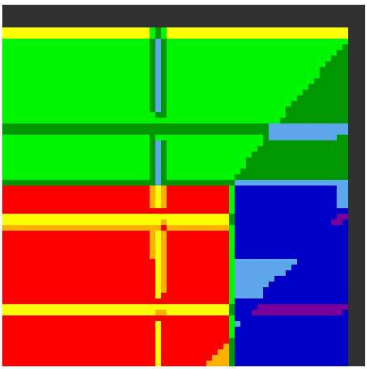}
\caption{PICsIT extended PIF model for science window 022000970010 calculated for 
Crab seen at 36$\degr$ off-axis angle and 53$\degr$ azimuthal angle. The mean wall 
transparency for 450 keV photons is 63\% for the lower left part (red), 39\% for 
the upper part (green) and 20\% for the lower right part of the shadowgram 
(blue).}
\label{offpifs}
\end{figure}

High energy photons from the source can pass through the IBIS tube walls 
and illuminate the part of the detector that is not coded by the IBIS mask. For 
that reason the non-coded part of shadowgrams cannot be used to estimate the 
background level and the PIF is usually calculated only for the coded part. 
However, the ISGRI PIF model implemented in OSA takes into account different 
structures such as walls or hopper, providing a source illumination pattern for 
the entire detector plane. This PIF model can be adapted for PICsIT after 
applying shifts and changing the size of the pixels. Tube walls are quite 
transparent to photons in the PICsIT energy range, for example, for sources at an 
off-axis angle of 36$\degr$ and with an azimuthal angle of 53$\degr$ about 40\% 
of 450 keV photons can pass through the wall, illuminating the detector surface. 
Therefore, it appears desirable to use such data in attempting to increase the 
amount of information about a given source where the large off-axis angle 
observations correspond to the exposure time, which can be several times longer 
than the time of fully coded observations. Figure \ref{offpifs} shows an example 
of the extended PIF model for the source observed at 36$\degr$ off-axis. The 
extended PIF model for PICsIT spectral extraction was tested with the Crab data, 
and the results of these tests are presented in Sect. \ref{other}. 

\section{Background model}

Given the fact that the PICsIT instrumental background rate exceeds the source 
count rate by several orders of magnitude, a correct analysis of PICsIT data must 
be based on careful modelling of the background. The background model for a given 
observation must be prepared after detailed studies of general PICsIT background 
properties and investigation of the behaviour of the background during that 
observation. Positions of the $\gamma$-ray lines are one of the basic 
characteristics of the background. Since these lines are emitted by isotopes 
produced by cosmic rays in the detector and the surrounding materials, a spatial 
distribution of the count rate over the detector surface is energy dependent. 
Another important property of the background is its time evolution. Long-term 
variability should be monitored to check, for example, the positions of the lines 
in the background spectra. Short-term variability must be studied for selection 
of data that can be used to prepare the background map for a given observation. 
Solar flare periods or time intervals with a strange background behaviour should 
also be excluded. This section presents the results of the PICsIT background 
studies.

\subsection{Spectra}

\begin{figure}[!h]
\centering
\includegraphics[width=9cm]{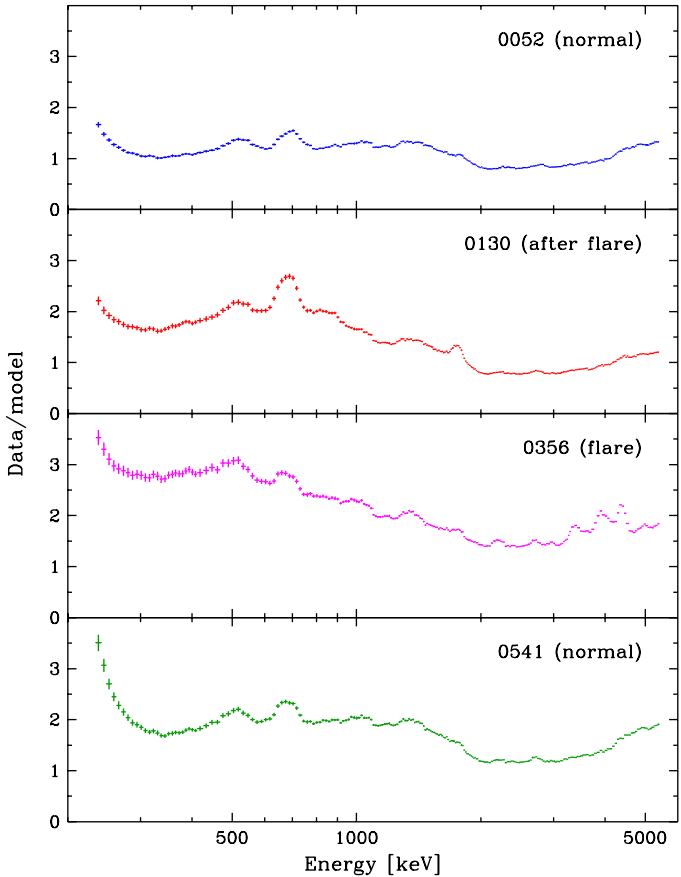}
\caption{PICsIT background spectra from Revs. 0052, 0130, 0356, and 0541. Data are 
divided by the power-law model with a spectral index $\Gamma$ set to 1 and 
a normalization at 1 keV equal to 1.25 keV$^{-1}$ cm$^{-2}$ s$^{-1}$. Spectra 
representing non-flaring Sun periods (0052, 0541) are separated by exactly 4 
years. The spectrum from Rev. 0130 was taken 3--5 days after the most powerful 
solar flare ever recorded (X28+, 4 Nov 2003, \citealt{Space07}), when the Sun 
already returned to quiescence. The Rev. 0356 spectrum shows the background 
emission observed 5--7 days after the fifth largest solar flare (X17, 7 Sep 
2005, \citealt{Space07}), while there was still considerable flaring activity.}
\label{back}
\end{figure}

Figure \ref{back} shows PICsIT background spectra from different periods of the
mission.  The spectra were integrated over one spacecraft revolution (hereafter 
Rev.) lasting almost 72 hours.  The spectrum of Rev. 0541 (19--21 March 2007), 
taken exactly 4 years after the Rev. 0052 spectrum (18--20 March 2003) 
illustrates how the background emission evolves during the mission. The sharpest 
increase in the emission with time was observed at low energy, below 300 keV, 
where the Rev. 0541 count rate was about as twice high as the Rev. 0052 count 
rate. Above 300 keV, the count rate after 4 years was about 50\% higher than at 
the beginning of INTEGRAL operation.

The PICsIT energy resolution is rather  coarse, varying between 30\% at 300 
keV and 4\% at 4 MeV \citep{Malaguti03a}. In spite of this, the background 
spectra are not completely smooth, exhibiting strong peaks at 511 and 680 
keV, weaker peaks at 1800 and 2750 keV, and a broad hump between 800 and 1700 
keV. The identification of the background characteristic features can be useful 
in properly adjusting the energy bins used in the data analysis and monitoring 
the energy calibration stability. Due to the limited energy resolution even the 
strongest lines positions were affected by contamination from weaker lines 
nearby, while weaker lines were hidden in the Compton continuum associated with 
the stronger lines, which form broad humps everywhere below 1700 keV. 
Therefore, only several background lines could be identified with a high level of 
plausibility. These identifications were supported by comparison with the list of 
SPI background lines \citep{Weidenspointner03} that were clearly resolved, and 
the data about CsI proton activation measurements \citep{Ruiz94}. The main lines 
identified in normal (non-flaring Sun periods) 
spectra are listed in Table \ref{lines}. The two strongest lines, the 511 keV 
($\beta^{+}$ decay line) and the 680 keV complex, are emitted predominantly by 
the iodine, caesium, and antimony isotopes produced in the detector material. 
Radioactive bismuth is produced in the BGO crystals of the IBIS anticoincidence 
system, whereas sodium and aluminum activity originates in the aluminum frame 
holding both IBIS detectors. 

In addition to the lines presented in Table \ref{lines}, normal spectra also 
exhibit weak emission around 2230 keV, presumably from a 
$^{1}$H(n,$\gamma$)$^{2}$D reaction. The bumps observed between 800 and 1700 keV 
are probably associated with emission from numerous isotopes produced mainly in 
the detector and its frame, for example $^{59}$Fe, $^{67}$Ni, $^{28}$Mg, and 
$^{120}$Sb. In the spectrum from Rev. 0356 taken during flare, the Sun and Earth 
atmosphere emission by light isotopes is observed at high energy. There are five 
lines at about 2200, 3400, 3900, 4150, and 4450 keV, which can be identified with 
the emission from ($^{15}$O, $^{11}$B, $^{10}$B, $^{16}$O, $^{15}$N, $^{14}$N, 
$^{13}$N), ($^{14}$N), ($^{13}$C, $^{14}$N), ($^{16}$O) and ($^{12}$C, $^{11}$B), 
respectively \citep{Share01}. During quiet Sun states, the Earth atmospheric 
emission is not observed, as demonstrated by analysing the spectra extracted from 
Revs. 0401, 0404, 0404, and 0406, when the INTEGRAL Earth occultation observation 
\citep{Churazov07} was performed. 

\begin{table}
\begin{center}
\caption{Strongest PICsIT background lines seen in the spectra collected during 
non-flaring Sun periods. The most probable line origin is listed first. 
T$_{1/2}$ is the half-life decay time of a given isotope. Lines around 670 and 
690 keV form the 680 keV complex (see text).}
\label{lines}
\begin{tabular}{cccc}
\hline\hline
Energy & Catalog energy & Isotope & T$_{1/2}$ \\
\multicolumn{1}{r}{[keV]} & [keV] & & \\
\hline
 510 &  511.0 & $e^{+}e^{-}$ & --- \\
 670 &  666.3 &  $^{126}$I & 13 d \\ 
     &  667.7 & $^{132}$Cs &  7 d \\
     &  666.3 & $^{126}$Sb & 12 d \\
 690 &  695.0 & $^{126}$Sb & 12 d \\
     &  685.7 & $^{127}$Sb &  4 d \\
     &  703.4 & $^{205}$Bi & 15 d \\
1780 & 1779.0 &  $^{28}$Al &  2 m \\
     & 1757.6 &  $^{57}$Ni & 36 h \\
     & 1771.4 &  $^{56}$Co & 77 d \\
2750 & 2754.0 &  $^{24}$Na & 15 h \\
\hline\hline 
\end{tabular}  
\end{center}
\end{table}

\subsection{Energy calibration}

A routine energy calibration for PICsIT is based on the special data (so-called
S5 datastream) collected in coincidence with the emission from the onboard 
$^{22}$Na calibration source \citep{Bird03}. S5 data are collected as separate 
64-channel spectra and monitored by the instrument team \citep{Malaguti03a}, 
providing the parameters for the channel-energy conversion. This standard 
calibration assumes that energy is a linear function of the channel number, which 
in general is not necessarily true. Studies of 256-channel background spectra 
completed directly by using the standard spectral-imaging histograms of single 
events can help in controlling the energy calibration, as well as extending it 
outside the 511--1274 keV range of S5 data.

Energy calibration tests were completed as follows. Background spectra were 
extracted for all public data from revolutions 0039--0676. The positions of the 
511, 667, and 1779 keV lines were then fitted in channel space using Xspec 11 
\citep{Arnaud96}. The lines were modelled with Gaussian shape and the local 
continuum was asummed to be in the form of a powerlaw.  After some initial 
calibration tests, the most reliable gain value was established to equal the 
width of one basic channel, i.e., 7.1 keV, in agreement with the standard energy 
calibration performed by the IBIS team \citep{Foschini07b}. The only remaining
parameter needing to be adjusted was then an offset, a shift in energy units to
be applied to the original onboard histograms to ensure that the three tested 
lines were in the correct positions. The best-fit offset was found to be 
+9.6 keV with respect to the standard energy calibration of PICsIT. The first 
usable channel of spectral-imaging single-events histograms was channel number 
10, which in the standard OSA 7.0 calibration starts at 203 keV and in the new 
calibration presented here at 212.6 keV. 

The mean positions and 1-$\sigma$ errors found for the background  lines in the
spectra from revolutions 0170--0676 are equal to 509.2$\pm$5.0 keV, 668.5$\pm$3.8 
keV and 1772.8$\pm$4.1 keV. Before revolution 170, a different binning was 
applied to the onboard histograms and the data were treated separately. The 
corresponding mean positions of the calibration lines were found to be shifted by
about +5 keV with respect to the results quoted above. 

The main factor driving the offset of energy calibration is the temperature of 
the PICsIT detector \citep{Malaguti03a}. Short time variations in the line 
position reach as high as about 15 keV on timescales of weeks. This is a large 
value compared to the minimal width of the bin used in spectral extraction, 
21.3 keV. In principle, this can introduce fluctuations in the extracted count 
rates. As we show in Sect. \ref{osa}, Crab count rates are stable over the 
mission time, which indicates that the offset correction is not very important,
although it should be taken into account in a proper preparation of the 
background maps. This is currently achieved by preparing several maps 
for revolutions with large temperature variations, using data from periods with 
an approximately constant temperature.  

\subsection{Variability}
\label{bgvar}

\begin{figure}[!h]
\centering
\includegraphics[width=9cm]{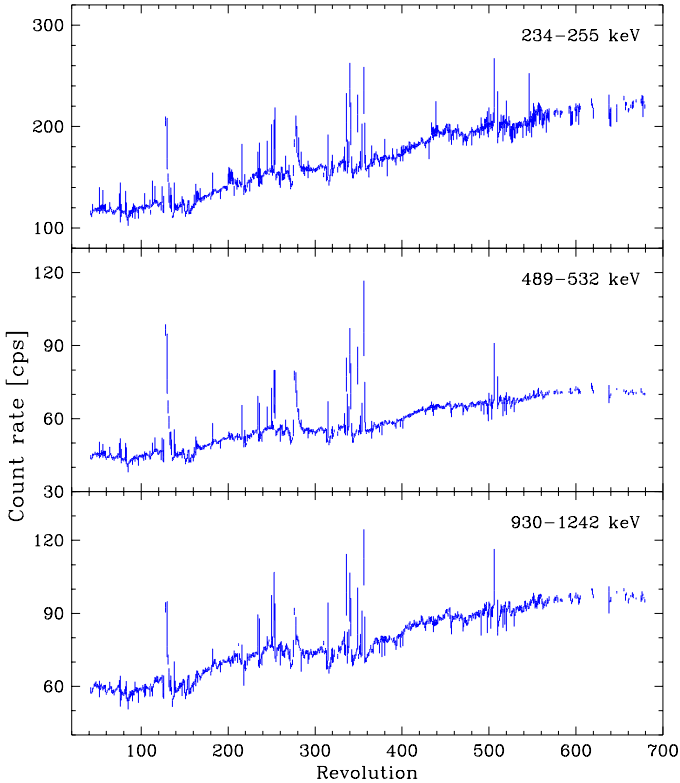}
\caption{Variation in the PICsIT background count rate during the mission.
Error bars correspond to the standard deviation of the mean for a given 
revolution and are much larger during solar flaring periods.}
\label{scarat}
\end{figure}

The count rate of the PICsIT background varies strongly in response to the 
cosmic-ray intensity changes, since it is moderated by the decay time of various 
radioactive nuclei produced in the detector neighborhood. On average, the 
background count rate increased with time during the first four and a half years 
of INTEGRAL operation, as shown in Fig. \ref{scarat}.  This increase was caused 
by both the solar modulation of the cosmic-ray intensity and the accumulation of 
the radioactivity in the detector assembly. The increase in the background count 
rate is irregular, there are periods of high variations but there are also 
periods of almost  constant count rate. Since March 2007 (around Rev. 0540),
the background appeared to have saturated at a level of about 60\% higher than 
the initial one. However, below 250 keV the increase in the background count rate 
is far more significant, higher than 100\% since the beginning of the mission and 
appears still not to be saturated.  

\begin{figure}[!h]
\centering
\includegraphics[width=9cm]{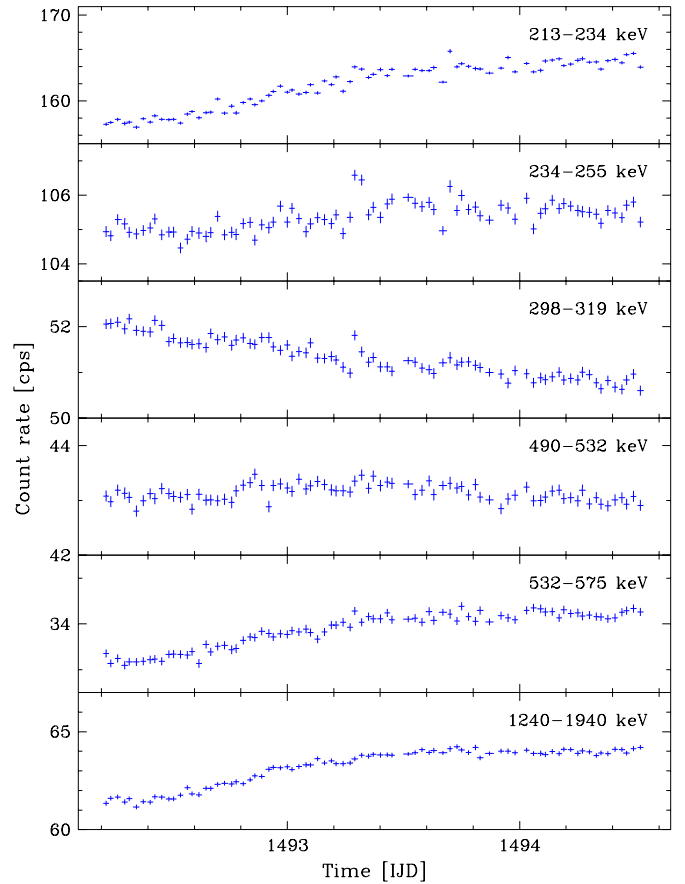}
\caption{Total PICsIT count rate measured in eight energy bands during Rev. 
0159. IJD is the INTEGRAL Julian Date (IJD = MJD - 51544.0).}
\label{revrat}
\end{figure}

Within a single revolution, a substantial variability is also observed, induced 
mainly by the spacecraft passages through the Earth radiation belts and the Solar 
activity. An example of this variability is presented in Fig. \ref{revrat}, for 
six energy bands between 213 keV and 1940 keV. This figure illustrates a rather 
typical behaviour, where the total count rate decreases with time for energies 
below $\approx$511 keV and increases above that energy. The very low energy bands 
(close to 200 keV) usually exhibit a different type of variation, due to the 
track events. The second and third (from top) panels of Fig. \ref{revrat} 
indicate that the background variability can be quite different in two close 
energy bands. Consequently, spectral extraction with a background model 
constructed for too wide an energy band can be inefficient when the radiation in 
the low-energy part of the band increases but in the high-energy part decreases 
with time. In general, there is a variety of background light curves observed, 
sometimes the changes in all energy bands are correlated, sometimes there are 
flare-like events or rapid count-rate variations that are only observable for a 
limited energy range. Therefore, for a correct background modelling, the
variability in narrow energy bands must be determined for each revolution.

\subsection{Background maps}
\label{maps}

The background model for PICsIT is the map of pixel count rates corresponding to 
the background emission registered in a given energy band during some period of 
time. Both the sum of photons originating in all point sources in the field of 
view and the summed diffuse sky emission equal at most of the order of several 
per cent of the total detector count rate at energies above 200 keV. Accordingly, 
the background map consists of a sum of shadowgrams collected during a number of 
dithering observations, such that the point sources and the diffuse emission 
patterns are spread over the detector surface. In this way, the background model 
contains the same data used subsequently to extract the source count rate. The 
dithering observational strategy thus allows us to obtain for the coded-mask 
instruments the highest quality background estimate, based just on the actual 
data. This is one of the most important virtues of masked pixellated detectors 
compared with standard telescopes, where the background must be determined in 
some indirect way, such as a rocking observing strategy as applied for OSSE.

While there is no variation in count rate within each pixel during an 
observation, the background map provides the most accurate, 
experimentally-determined background model, with an accuracy increasing with the 
exposure time. This situation never occurs in reality and the main task in 
achieving a correct background model is to describe properly its variation in 
time. Some observations are also taken in special conditions, such as staring 
mode, when the map cannot be prepared with the contemporary data. Consequently, 
the variability model should allow us to check if another 
map can be used for these special observations. The simplest way of taking into 
account the background variability in time is to assume that there is only one 
global trend, coherent for all pixels. The only parameter that has then to be 
determined with the source count rate is the background map normalization for a 
given science window. As shown in Sect. \ref{spectra}, this approach produces 
completely acceptable results. 
 
In view of the considerations mentioned above, the only parameter that is free 
in the modelling of the background is the time period for which the background 
map is prepared. Since the spacecraft activation is caused predominantly by 
passages through the Earth radiation belts, the natural choice of time interval 
for which a map is prepared is the revolution period. Owing to the large number 
of counts collected by each pixel during this time period, the accuracy of the 
revolution-averaged background rate is sufficiently high. On the other hand, if 
there is a sub-orbit variation observed, the shortest time for which the 
background map can be prepared is limited by the condition of smoothing 
the source illumination pattern. This will depend on the actual dithering 
pattern, but usually about 10 science windows are sufficient to obtain a 
sufficiently smoothed summed shadowgram. Nonetheless, producing more than one 
background map for a single revolution is unnecessary, unless there is clear 
variability in the background mean count rate or the detector temperature 
measured during the orbit time. In practice, almost all results presented later 
were obtained with the use of a single map, apart from several observations 
of Crab and Cyg X-1, where two maps per revolution were needed to suppress an 
artificial trend found in the source light curves.

Data used to prepare the map were selected according to the mean count-rate 
level over all pixels in a given science window. The fluctuations in that value 
usually did not exceed 3\% for periods with a stable background and this 
criterion was adopted for the PICsIT data analysis presented in this paper. 
In the spectral extraction, a similar criterion was used, excluding the data,  
when the background map normalization was outside the 0.97--1.03 range. This 
selection automatically excludes periods with solar flares or science windows 
affected by the passage through the radiation belts at the beginning or end of 
revolution. 

\begin{figure}[!h]
\centering
\includegraphics[width=9cm]{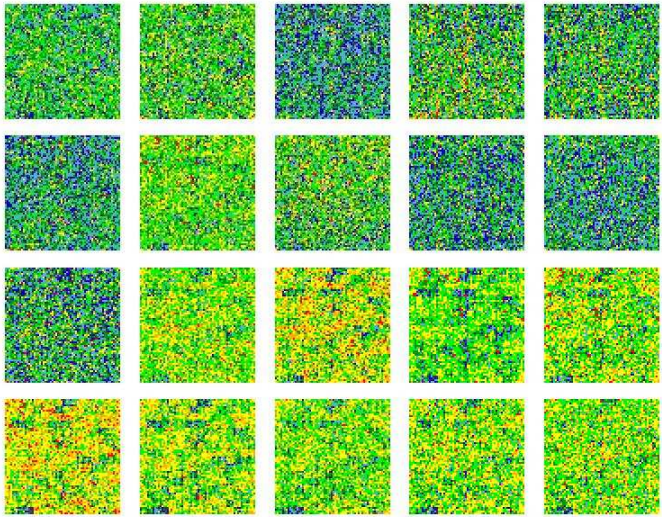}
\caption{Distributions of differences between 336-448 keV shadowgram from Rev. 
0079 and corresponding shadowgrams from Revs. 0070 to 0078 and 0080 to 0090 
(starting from the upper left, row by row).  Warmer colours (yellow, red) 
correspond to the higher count rates.  The smallest differences (flattest 
images) are observed for Revs. 0072, 0075, 0078, 0080 and 0081. There was a 
small solar flare in Rev. 0076 that changed also the background distribution 
in the following revolution.}
\label{diff79}
\end{figure}
 
When it was not possible to prepare the map for part or an entire revolution, 
one could use, in principle, a map prepared for another period close in 
time. Tests done for the Crab indicated that this approach did not always 
provide good results, especially when the map is constructed with data from 
another revolution. The simplest way of confirming the adequacy of a map is an 
investigation of the differences or ratios between the pixel count rates in the 
summed shadowgrams from both the map and the source observation. Figure 
\ref{diff79} presents an example of this test made for the summed shadowgram from 
Rev. 0079 when Cyg X-1 was observed. The flattest difference shadowgrams are 
those for Revs. 0078 and 0080 (two rightmost in the second row of the plot). It 
has to be kept in mind, however, that this test may not be applicable when a 
relatively strong source is observed in staring mode, because then there will be 
a systematic imprint on the shadowgram from the source illuminating the same 
pixels all of the time. A possible solution to this limitation is to test the 
background stability over several revolutions adjacent to the actual source 
observation and to prepare some mean map.

\section{Count-rate extraction method}
\label{ppdf}

The necessity of merging a large amount of PICsIT data to detect weaker sources
not only compels one to model the background precisely but also to apply a method 
that is strictly correct for data with a low signal-to-noise ratio. Otherwise, 
the  inadequacy  of the method can produce results with a bias comparable to the
signal strength. The most common way of estimating some parameter value and its 
confidence limits is the application of the $\chi^{2}$ test statistic. Although 
it is well known that the conditions needed to justify the use of the $\chi^{2}$ 
test are not fulfilled for low number of counts, it is worth reminding us of all 
limitations of the standard technique. 

1. The assumption that the Poisson distribution can be approximated by a 
Gaussian distribution for a large number of counts is not always justified 
because these distributions differ: the Poisson distribution is asymmetric 
with non-zero skewness and kurtosis, whereas the Gaussian distribution is 
symmetric with zero skewness and kurtosis. When the signal has an 
amplitude comparable with the difference in shapes of the Poisson and 
Gaussian distributions for the total (signal+background) number of counts, 
the Gaussian approximation will provide a systematically biased result. 

2. The Poisson distribution is defined for parameters of non-negative values.
Count rates also cannot be negative and any analysis using a Gaussian or other
distribution in negative signal space does not offer a completely proper 
description of the phenomenon. 

3. The difference of two parameters both characterized by a Poisson 
distribution is not Poisson distributed. Consequently, the signal+background and 
the background alone should be modelled simultaneously instead of 
subtracting the background rate from the total rate, especially when this 
subtraction leads to negative signals.

4. Since the true variance is usually unknown, the observed number of counts 
serves as a variance estimate for the $\chi^{2}$ statistic. This approximation
is invalid for a small number of counts and many solutions have been proposed to 
improve the variance estimate, but none of them performs well for weak signals, 
when zero net counts are measured in a majority of trials.

5. To estimate the uncertainty of a parameter of interest when many 
parameters are fitted simultaneously, one usually considers the limits 
corresponding to the change in the statistic by a certain value 
\citep{Lampton76}. This approach based on the projection of the respective
parameter confidence region onto the parameter axis often overestimates the 
error. Moreover, its basic condition, namely the use of the model variance, is 
not met in many practical applications, as stated in point 4.

The solution to several issues listed above is application of a statistic 
proposed by \citet{Cash79} based on a maximum likelihood estimation for the 
Poisson distribution. The application of the Cash statistic allows us to avoid 
problems arising due to the differences between Poisson and normal distributions 
but still has some limitations that will be discussed in Appendix A. Due to its 
direct handling of probability density function (hereafter PDF) associated with 
the model parameters, a Bayesian approach provides a more general and efficient 
solution to many statistical questions, including the `low-number-of-counts' 
issue. 

A Bayesian technique for dealing with the Poissonian type data in the presence of 
known or unknown background was presented by \citet{Loredo90}. This type of 
approach applied to a low-quality ASCA data obtained for weak AGNs, was 
demonstrated to reconstruct precisely the assumed shape of the iron line complex 
\citep{Lubinski04}. Due to its efficiency, the method has now been generalized to 
deal with the extraction of count rates for objects observed with PICsIT. 
Developments have occured in two respects: the Poisson PDF is now computed in two 
(or more) dimensions, and the model now takes into account the main observation 
conditions.

The probability density function $p_{k}(s,b_{k})$ of the source count rate 
parameter $s$ and the background map normalization factor $b_{k}$ in the science 
window number $k$  for a given energy band  is defined as the product of Poisson 
PDFs $P$ for all $n_{k}$ active pixels 

\begin{equation}
    p_{k}(s,b_{k}) = \frac{1}{C_{k}}\prod_{i=1}^{n_{k}}P(N_{k}(i);
    (s\eta_{k}\xi_{k}(i)+b_{k}B(i))f_{k}(i)),
\label{pdfw}    
\end{equation}

\noindent where the factor $C_{k}$ normalizes the $p_{k}(s,b_{k})$ integral to 1, 
$N_{k}(i)$ is the number of counts measured by the $i$-th pixel, $\eta_{k}(i)$ is 
the source PIF value for that pixel, $\xi_{k}$ corrects $s$ for the off-axis 
effect, and $B(i)$ is the efficiency-corrected background count rate given by the 
background map. The factor $f_{k}(i)$ converts count rates into counts,

\begin{equation}
    f_{k}(i) = T_{k}\varepsilon_{k}(i)/(32\times64),
\end{equation}

\noindent using the exposure time $T_{k}$ and the pixel efficiency 
$\varepsilon_{k}(i)$, where the source count rate is normalized to the half of the 
detector area (32$\times$64 pixels).

The density $p_{k}(s,b_{k})$ is the posterior probability density, i.e. the prior 
probability densities on $s$ and $b_{k}$ are assumed to be uniform. When a large 
amount of data is used, as in the PICsIT case, with thousands of pixels and 
hundreds of science windows, the final density distribution associated with the 
detected source signal is sufficiently concentrated not to be affected by the 
choice of the prior distribution. 

Since the calculation of the PDF must be performed for a relatively wide 
range of $s$ and $b_{k}$, the product in Eq. \ref{pdfw} is computed by summing 
the PDF logarithm to avoid the multiplication of many small numbers. The final 
PDF $p(s)$ is derived after integrating $p_{k}(s,b_{k})$ over the background 
normalization parameter  and then computing the product of $p_{k}(s)$ for all
science windows of interest. Using $p(s)$, one can extract all needed parameters 
such as the mean, the median, or the desired credibility intervals. For PICsIT, 
it is possible to have a detection in a single science window only for sources 
at least as bright as the Crab. To achieve a reliable result for weaker 
objects, a large number of science windows must be merged. Table \ref{counts} 
compares typical values of the number of counts, count rates, and signal-to-noise 
ratios for data of a 50 mCrab source observed with PICsIT and a 2 mCrab source 
observed with ISGRI. 

\begin{table}
\begin{center}
\caption{Typical source ($R_{s}$) and background ($R_{b}$) count rates observed 
with PICsIT (upper part, Rev. 0301) and ISGRI (lower part, Rev. 0079) for sources 
of 50 mCrab and 2 mCrab strength, respectively. Corresponding numbers of counts, 
$N_{s}$ and $N_{b}$, were calculated for science windows with a 2 ks exposure 
time. S/N is the signal-to-noise ratio, and $k_{3\sigma}$ is the number of 
science windows with the 2 ks exposure time needed to obtain a 3-sigma detection.
}
\label{counts}
\begin{tabular}{ccccccc}
\hline\hline
Energy range & $R_{s}$ & $N_{s}$ & $R_{b}$ & $N_{b}$ & S/N & $k_{3\sigma}$ \\
\multicolumn{1}{c}{[keV]} & [cps] & & [cps] &  &  & \\
\hline
 298-319 & 0.046 & 74 & 52.0 & 104000 & 0.23 & 170 \\
 433-461 & 0.017 & 24 & 32.0 & 64000 & 0.10 & 900 \\
 717-816 & 0.010 & 15 & 36.0 & 72000 & 0.06 & 2500 \\
\hline
 22-30 & 0.184 & 294 & 36.0 & 57600 & 1.22 & 6 \\
 71-80 & 0.019 & 30 & 22.4 & 35840 & 0.16 & 350 \\
 150-250 & 0.008 & 12 & 53.6 & 86400 & 0.04 & 5625 \\
\hline\hline 
\end{tabular}  
\end{center}
\end{table}

\begin{figure}[!h]
\centering
\includegraphics[width=9.2cm]{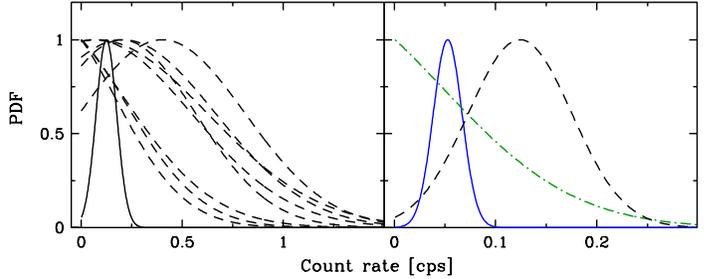}
\caption{Examples of Poisson PDFs obtained for GRS 1758-258 in the 298--319 keV 
band. Left: broad distributions obtained for several single science windows from 
Rev. 0105 (dashed lines) and a narrower PDF from all Rev. 0105 data merged (solid 
line). Right: The same total PDF from Rev. 0105 (dashed line) compared to that of 
Rev. 0103 (dot-dashed line), and the final result for all data of August -- 
October 2003 period (solid line), with an exposure time of 1.95 Ms.}
\label{pdfs}
\end{figure}

Examples of Poisson PDFs obtained for a true source are presented in Fig. 
\ref{pdfs}. PDFs computed for a single science window are very broad when 
compared to the source count rate. For an object as weak as GRS 1758-258 (about 
70 mCrab in the 300--500 keV band), even one-revolution of data is not enough to 
achieve a detection. Nevertheless, thanks to a frequent monitoring of the 
Galactic centre field by INTEGRAL, it is possible to follow the seasonal (i.e. 
on the scale of several months) changes in the high energy emission from this 
source in several energy bands below 500 keV.

The technique presented in this paper is the most correct means of handling the 
Poisson-distributed data within the Bayesian framework. Due to this first 
application of such a method to extract the astrophysical $\gamma$-ray data, 
some tests of the performance are needed to check for example the limiting signal 
that can be extracted and compare this technique with several other commonly used 
methods. An obvious test is to use this method for the ISGRI spectral extraction 
and to compare the results with those of the standard OSA software. The results 
of these tests are presented in Appendix A. 

The computation time needed to extract the source count rate over a PDF grid as 
fine as possible (the grid step should be much smaller than the expected rate) is 
long compared to the time needed by the method using the fitting with some 
statistic. For only two parameters handled (the count rate for a single source 
and the background normalization), computing over a PDF grid can be 
performed even for a large data set. However, if there are several sources in the 
field of view, the computation over the grid has to be replaced by some Monte 
Carlo integration of multi-dimensional PDFs. Such procedures are time consuming 
and need some adjustments to accelerate the data analysis. Due to the small 
number of sources observed with PICsIT, this type of software does not appear to
be needed, especially in a situation when the emission from any contaminating 
source is far weaker than the background. Nevertheless, a version of the software 
with the Monte Carlo integration implemented was prepared and tested with the 
simulated data for up to 7 sources in the field of view. These tests show that 
only in the case of sources separated by a short distance (less than several 
degrees) there can be some contamination for weaker sources. An independent test, 
where the single-source PDF method was applied separately to ISGRI data of NGC 
4151 and NGC 4051 (separated by 5.3$\degr$), has shown that the results for both 
objects are fully consistent with those of the standard OSA software handling 
many sources at once. 

\section{Detection limits}
\label{limits}

The standard way of verifying whether the observed signal corresponds to a 
detection is the `n-sigma' significance test. The test statistics 
(significance) is defined as the difference between the measured 
signal+background and the background estimated in an independent way, divided by 
the noise represented by the standard deviation of that difference. Detection can 
be claimed if the observed significance is higher than the threshold for a given 
probability level, e.g., 3 for probability level 0.997 (3-sigma test). The 
significance level is often set to 5- or 6-sigma because such a conservative 
approach is supposed to account for the number of trials or to balance all 
possible systematic effects not incorporated into the uncertainty calculation. 

In true situations, the significance estimate should take into account all 
circumstances of the measurement: the detector model, the background model, and 
the method used to extract the count rate. For coded-mask instruments, one has to 
take into account the mask pattern and transparency, the spatial resolution of 
the detector and the background non-uniformity. There are many approaches 
developed to determine analytically the sensitivity of coded-mask instruments 
(see \citealt{Skinner08} and references therein). 
 
This simple `n-sigma' significance test cannot be applied to the results of the
technique presented here. The method based on the Poisson PDFs extracts the net 
source count rates without direct background subtraction and operates in a 
physical, non-negative parameter space. The mean net count rate determined by 
integrating a PDF over a non-negative argument cannot be zero even when there is 
no signal, although it decreases with increasing exposure time. Therefore, there 
will always be some signal excess above zero resulting in an overestimation of 
the true significance level.  On the other hand, for the weak signals expected to 
be observed with PICsIT, a too conservative approach might exclude some potential 
sources from the analysis. Therefore, we would like to find a way of estimating 
the true noise level. This will be useful not only for the verification of the 
results but will also help to judge if the source with a given flux level 
can be detected after a given observation time.

The question addressed here is: ``what is the probability of obtaining a false 
signal, arising from the background (noise) fluctuation and described by the 
probability density function with a given mean count rate, for an observation 
lasting a given time". To find the answer, we compiled the distribution of mean 
count rates extracted for random positions in fields that do not include strong
sources. The tail of the distribution was then fitted to find the upper noise 
limit corresponding to a given probability level. Tests of the PICsIT noise level 
were completed for six wide energy bands: 277--362, 362--461, 461--632, 632--930, 
930--1938, and 1938--3131 keV.  The limits of these wide bands were adjusted to 
have a coherent background variation within a given band (see Sect. 
\ref{bgvar}). Input empty-field data were selected as observations when all of 
the eight possibly brightest PICsIT sources (Crab, Cyg X-1, GRS 1758-258, 4U 
1700-377, GRS 1915+105, Cen A, 1E 1740.7-294, XTE J1550-564) were farther than 
45$\degr$ away from the pointing direction. This criterion provided 5599 science 
windows from Revs. 0047--0450 with the total exposure time equal to 14.48 Ms. 
Input PIF data were calculated for 91 (13$\times$7) possible positions 
corresponding to the hexagonal dithering pattern with a so-called ``wandering 
centre of pattern" \citep{Kuulkers05}. The hexagonal dithering pattern was chosen 
because in this case the object is in the fully coded FOV all the time and there 
is no need to correct the exposure time for incomplete detector illumination. 

At the beginning of each test, the PIF was randomly selected for each input
science window. The net source count rate was then extracted for each science
window using the method described in the previous section. These results were
merged for a  subset  of all science windows to obtain the mean count rate 
for a given assumed total exposure time. For shorter times, it was possible to 
select the subset several times because the entire sample was sufficiently large 
to avoid repetition. To ensure that the tail of the mean count-rate 
distributions could be determined to a good precision, approximately 1000 tests 
had to be performed. However, since computation of one test takes several hours 
on a 20-CPU grid network, it is impossible in practice to complete 1000 tests 
for times longer than 1 Ms. 

Upper noise limits were calculated for probability values corresponding to the 
commonly used `n-sigma' significance levels, where n was in the range 1--5. The 
limit of 1-sigma significance level was determined as the mean 
count rate value for which the sum of runs, starting from the minimal mean count 
rate, reaches 68.3 \% of the total run number. A model function was then fitted 
to the tail of the distribution, i.e. above a 1-sigma limit and the 2--5 sigma 
limits are found via integration of that tail model. Because the true shape of 
the distribution was unknown, three model distributions were tested for the 
277-362 keV band, where the number of tests was much larger, reaching more than 
11 000 runs for the shortest 10 ks exposure time. The Gaussian function was used 
to model a rapid decline and the exponential function to model a slow decrease. 
The third model included a gamma distribution, which reproduced an intermediate 
decrease. For distributions analyzed with a large number of tests ($>$ 1000), the 
best fit was always obtained with the Gaussian distribution. The fitted gamma 
distribution was usually quite similar to that of the Gaussian, but the 3-sigma 
upper limit was found to be slightly (1-5\%) larger. The exponential model, in 
turn, seems to underestimate the decrease in the tail, giving always the highest 
$\chi^{2}$. The uncertainty in the fitted 3-sigma upper limit was estimated to 
be, in general, not larger than 10\% and 20 \% for the exposure times below and 
above 1 Ms, respectively. This estimate was based on tests with an increasing 
number of runs using different tail models.   

\begin{figure}[!h]
\centering
\includegraphics[width=9cm]{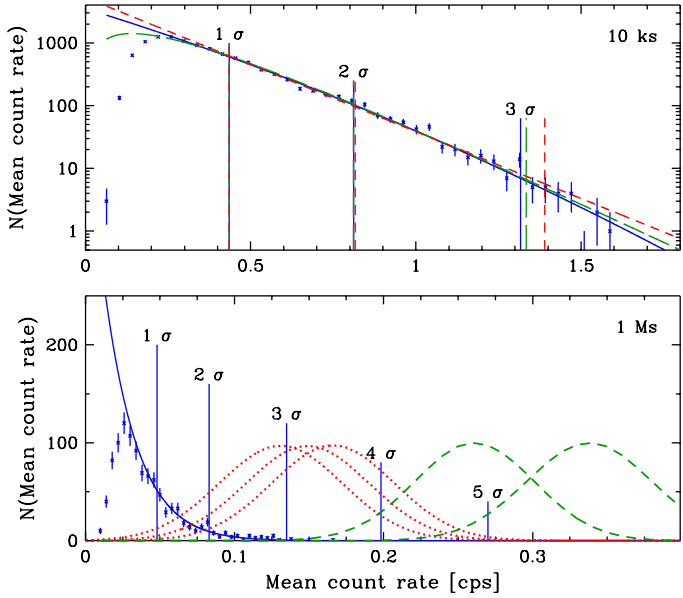}
\caption{Distributions of mean count rates in the 277--362 keV band extracted for 
fake sources in empty fields. Upper panel: Total exposure time 10 ks, 11 000 
tests. Solid, long-dashed and short-dashed lines show the Gaussian, gamma, and 
exponential models, respectively, fitted to the tail of the distribution, i.e. 
above the count rate corresponding to a 1-sigma upper limit. Vertical lines 
indicate the positions of the best-fit 2 and 3-sigma upper limits. Lower panel: 
Total exposure time 1 Ms, 1000 tests. The solid line shows the Gaussian function 
fitted to the tail of the distribution. Vertical lines present the 1 to 5 
sigma upper noise limits computed using the fitted Gaussian tail. Three 
extremely high-noise distributions for 1 Ms are shown with dotted lines, and two 
examples of count rate distributions obtained for GRS 1758-258 with a similar 
observing time are drawn with dashed lines.}
\label{estino}
\end{figure}

Figure \ref{estino} presents the example of noise limit estimate for the 277--362 
keV band and exposure times of 10 ks and 1 Ms. The distributions of mean count 
rates extracted for empty fields are asymmetric and clearly peaked above zero. 
Comparison of the three fitted tail models is shown in the upper panel of the 
figure. For 1 Ms, three PDFs are presented with the highest mean count rate, 
which illustrate the typical shape of the probability density distribution at 
this count rate level. These PDFs are compared with two PDFs obtained for the 
microquasar GRS 1758-258 after observations lasting about 1.05 Ms. The GRS 
1785-258 PDFs correspond roughly to 5 and 6-sigma detections.

\begin{figure}[!h]
\centering
\includegraphics[width=9cm]{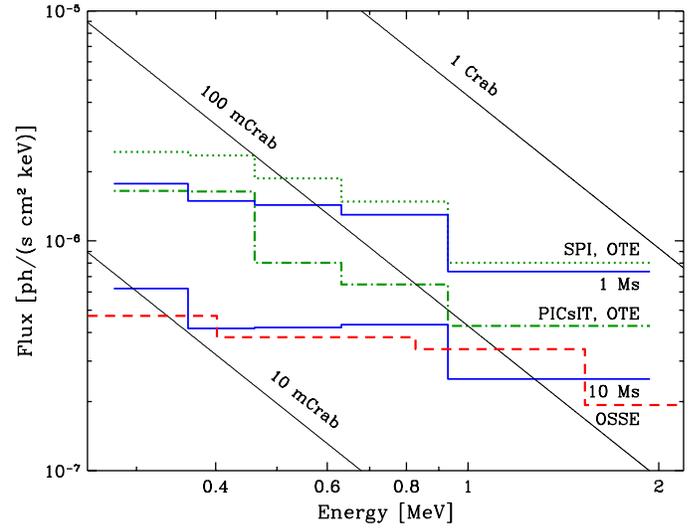}
\caption{PICsIT sensitivity limits for a 3-sigma detection for an exposure time
of 1 Ms. The thick solid lines show the estimates based on the noise level test 
for 1 and 10 Ms effective exposure times. Dotted and dash-dotted lines correspond 
to the OTE tool results for 1 Ms observation with SPI and PICsIT, respectively. 
The dashed line shows the 1 Ms sensitivity limits for OSSE. Thin 
lines show the fluxes for 10 mCrab, 100 mCrab and 1 Crab sources with the 
spectral slope fixed to 2.2.}
\label{sensi}
\end{figure}

The results of the noise level studies for 3-sigma detection limits are collected 
in Table \ref{limits1}. A wide range of effective exposure times should allow for 
a relatively precise interpolation for a particular observation time. The last 
row of the table shows the mean PICsIT fluxes measured for the Crab that can be 
used to convert the limits into Crab units. PICsIT sensitivity limits for 1 and 
10 Ms are compared in Fig. \ref{sensi} with the 1 Ms sensitivity limits computed 
in an analytical way for PICsIT, SPI, and OSSE. Sensitivity limits for PICsIT and 
SPI were calculated using the on-line Observation Time Estimator (OTE)\footnote{ 
http://integral.esa.int}. The formulae used by the OTE were described in 
\citet{Belanger08}. OSSE sensitivity data are taken from Fig. 3 of 
\citet{Winkler94}.

\addtolength{\tabcolsep}{-0.5mm}
\begin{table}
\begin{center}
\caption{PICsIT  3-sigma  sensitivity limits determined through the 
extraction of count rates for fake sources in the empty field observations. The 
last row presents mean fluxes determined for Crab.}
\label{limits1}
\begin{tabular}{cccccc}
\hline\hline
Eff. exp. & \multicolumn{5}{c}{Energy band [keV]} \\
& 277--362 & 362--461 & 461--632 & 632--930 & 930--1938 \\
\multicolumn{1}{c}{[ks]} & \multicolumn{5}{c}{[10$^{-6}$ ph cm$^{-2}$ s$^{-1}$ 
keV$^{-1}$]} \\
\hline
    10 & 16.3 & 15.3 & 11.2 & 13.4 & 6.99 \\
    20 & 11.1 & 10.2 & 8.58 & 9.11 & 4.94 \\
    50 & 6.49 & 7.37 & 6.39 & 6.18 & 2.85 \\
   100 & 5.25 & 4.97 & 4.35 & 4.27 & 2.24 \\
   200 & 3.92 & 3.63 & 3.46 & 3.01 & 1.59 \\
   500 & 2.47 & 2.44 & 2.37 & 1.55 & 1.18 \\
  1000 & 1.77 & 1.49 & 1.43 & 1.30 & 0.74 \\
  2000 & 1.10 & 0.95 & 1.09 & 0.98 & 0.43 \\
  5000 & 0.71 & 0.64 & 0.76 & 0.48 & 0.30 \\
 10000 & 0.62 & 0.42 & 0.42 & 0.43 & 0.25 \\
 14500 & 0.42 & 0.43 & 0.47 & 0.32 & 0.19 \\
1 Crab & 46.6 & 24.6 & 15.1 & 8.00 & 1.93 \\
\hline\hline 
\end{tabular}  
\end{center}
\end{table}
\addtolength{\tabcolsep}{+0.5mm}

Despite the different approach used in the computation, PICsIT sensitivities 
based on the noise level tests and given by the OTE tool are consistent for the 
first two energy bands. Above 460 keV, the OTE estimates are well below those 
determined experimentally. Since noise determination based on the empty field 
observations takes into account several effects not included in the OTE model 
(e.g. background non-uniformity, background level evolution with time, and a 
realistic PIF model) one can expect OTE to overestimate the PICsIT sensitivity. 
The same concern can possibly be raised for the SPI sensitivity estimates made 
in an analytical way. The PICsIT detector should be more sensitive to continuum 
observations because of its larger area and volume, the higher quantum efficiency 
of CsI compared with Ge crystals in terms of registering photons, and higher 
number of detector and mask pixels allowing for a more robust disentanglement 
between the signal and the noise. Nevertheless, a more reliable comparison of the 
SPI and PICsIT sensitivity is not possible unless a noise study similar to that 
completed for PICsIT is also completed for SPI. On the other hand, this 
comparison is probably more secure for the OSSE detector because of the rocking 
strategy of observations with that instrument. PICsIT appears to be much less 
sensitive than OSSE. As can be seen in Fig. \ref{sensi}, PICsIT needs about 10 
times longer exposure time to reach the OSSE S/N ratio because its volume is 
three times smaller. This lower efficiency is partly compensated, however, by a 
larger fraction of observing time (because of a higher orbit). 

\section{PICsIT spectra}
\label{spectra}

We now review the PICsIT spectra extracted for various objects. Some of these 
spectra are used to verify PICsIT calibration, by comparing them with the spectra 
obtained from two other high-energy INTEGRAL detectors, SPI, and ISGRI. Potential 
problems that might arise due to systematic effects are checked using Crab 
spectra. The results of the spectral extraction method based on the Poisson PDF 
(hereafter PPDF) technique described in Sect. \ref{ppdf} are compared with the 
results obtained with the standard OSA 7.0 software. Finally, examples of spectra 
derived for  several  detected objects are shown against the noise level expected 
for a given exposure time.

\subsection{Calibration}
\label{calibration}

Cross-calibration tests are one of the most important elements of the 
verification of a given detector performance. Due to the presence of three 
high-energy detectors onboard INTEGRAL observing the source at the same time, 
these tests are easier and more reliable than in the case of satellites hosting 
only a single instrument operating in a given energy range. Moreover, the SPI 
detector was carefully calibrated before the launch of INTEGRAL 
\citep{Attie2003}, which ensured that the calibration check was  
independent of the details of the particular model assumed for a calibration 
source such as the Crab. A lot of effort has been invested since the beginning of 
the mission to calibrate well the ISGRI and SPI instruments, resulting in a good 
performance in the energy range below 100 keV \citep{Jourdain08}. Due to a 
limited ISGRI sensitivity above 100 keV, there is however still a spectral slope 
issue remaining for the high energy part of the spectra. On the other hand, 
PICsIT standard OSA spectral extraction was applied only to a limited number of 
sources. This enabled only a crude check of the calibration, suggesting that 
there was an overall agreement between PICsIT and ISGRI spectral results 
\citep{Foschini07b}.

The first PICsIT response files for single-event spectra were provided to the 
user community with the OSA 4.2 release in December 2004. The two response files 
currently used are the RMF {\sl pics\_srmf\_grp\_0003.fits} (or {\sl 
pics\_srmf\_grp\_0005.fits} rebinned to the default energy bins) and the ARF 
{\sl pics\_sarf\_rsp\_0003.fits}. Both files were used in the tests presented 
here, although both interpolation and smoothing were applied to them to correct 
for discontinuities caused by the too wide energy bands used in their Monte Carlo 
computation. These corrections were necessary for obtaining smooth background 
spectra without jumps, observable when the standard responses are used (see e.g. 
Fig. 6 in \citealt{Belanger08}). The source spectra with finer binning are also 
smoother after this correction, although the overall calibration remains similar 
to the standard one. The final change that remains to be applied to the 
current response is the correction for the influence of the mask pattern. 
Preliminary tests showed that this correction affects mainly the high-energy part 
of the response, with the spectral fitting results below 1 MeV remaining 
virtually the same (L. Natalucci, private communication). 

Another issue related to the INTEGRAL cross-calibration is the ISGRI spectral 
slope above 100 keV. Due to superior sensitivity up to about 150 keV, ISGRI 
spectra dominate the INTEGRAL spectral fitting. Therefore, even a small 
discrepancy between the spectral slope of data from ISGRI and the two other 
detectors, can change substantially the results of a broad-band fit, producing a 
model with an incorrect spectral shape at high energy and incorrect relative 
normalization between instruments. For a long time, the photon index $\Gamma$ of 
the reference Crab model used in ISGRI calibration was set to be 2.225 for the 
entire ISGRI energy range. In OSA 7.0, following the recommendation of the 
INTEGRAL Users Group, it was changed to 2.1 below 100 keV and 2.34 above 100 keV, 
to match the SPI Crab spectrum, as provided by the SPI team. However, the latter 
value is clearly higher than $\Gamma$ $\approx$ 2.2 inferred from the Crab 
spectral analysis with the standard OSA SPI software, provided in its version 7.0 
and several earlier releases. The latest INTEGRAL cross-calibration report 
\citep{Jourdain08} presented a new result for SPI: with the updated spectral 
extraction software used by the SPI team, the Crab spectral slope above 100 keV 
is now equal to 2.22, i.e. a value fully consistent with the OSA result. One can 
thus expect the next OSA releases to provide revised ISGRI response files, 
adjusted accordingly. To anticipate this inevitable result, the standard ISGRI 
OSA 7.0 ARFs were corrected to match the value 2.22 above 100 keV for the purpose 
of the testing described below. 

\begin{table}
\begin{center}
\caption{Cross-calibration between the INTEGRAL high-energy instruments. A 
power-law model was fitted to the Crab (three upper rows) and Cen A (three lower 
rows) spectra. The best-fit parameters for Crab are: a photon index $\Gamma$ of 
2.19$\pm$0.01 and a normalization at 1 keV of 17.1$\pm$0.5 photons keV$^{-1}$ 
cm$^{-2}$ s$^{-1}$ with a reduced $\chi ^{2}$ of 1.71 for 49 degrees of freedom. 
The best-fit parameters for Cen A: a photon index $\Gamma$ of 1.80$\pm$0.01 and 
a normalization at 1 keV of 0.153$\pm$0.006 photons keV$^{-1}$ cm$^{-2}$ s$^{-1}$ 
with a reduced $\chi ^{2}$ of 1.16 for 61 degrees of freedom. No systematic 
errors were added to the statistical ones.}
\label{calib}
\begin{tabular}{cccc}
\hline\hline
Instrument & Exposure & Energy range & Rel. norm. \\
\multicolumn{1}{c}{} & [ks] & [keV] & \\
\hline
 ISGRI  & 1168 & 100--462 & 0.875$\pm$0.003 \\
 SPI    & 910 & 105--957 & (1.00) \\
 PICsIT & 981 & 277--930 & 0.75$\pm$0.01 \\
\hline
 ISGRI  & 771 & 19--462 & 0.89$\pm$0.02 \\
 SPI    & 584 & 26--352 & (1.00) \\
 PICsIT & 429 & 277-454 & 0.85$\pm$0.13 \\
\hline\hline 
\end{tabular}  
\end{center}
\end{table}

Calibration tests were made for two sources, \object{Crab} and \object{Cen A}, 
the latter to enable checks to be completed for a weaker object and with a 
different spectral shape. High energy spectra were produced with a large data 
set to achieve the highest possible precision. For Crab, all data from the 
standard dithering observations made in Revs. 0239, 0300, 0365, 0422, 0483, 0541, 
0605, 0665, and 0666 were used. This choice takes into account the fact that 
before Rev. 0216 SPI had not yet lost two of its detectors and its response was 
thus different during earlier observations. In the case of Cen A, it was 
impossible to obtain acceptable SPI spectra for observations completed between 
January 2005 and February 2007. Unfortunately, in this period 
there were strong background variations caused by enhanced Sun activity. PICsIT 
spectra for that period were also clearly worse than those obtained for earlier 
observations. The Cen A calibration spectra comprise therefore only data from 
observations made before 15 February 2004 (up to Rev. 0163), for which the source 
was detected at energies up to about 460 keV. Calibration tests with ISGRI and 
SPI spectra of the Crab were limited to the energy band above 100 keV, to 
decrease the influence of the low energy part on the fit completed for the 
PICsIT spectral band.

\begin{figure}[!h]
\centering
\includegraphics[width=9cm]{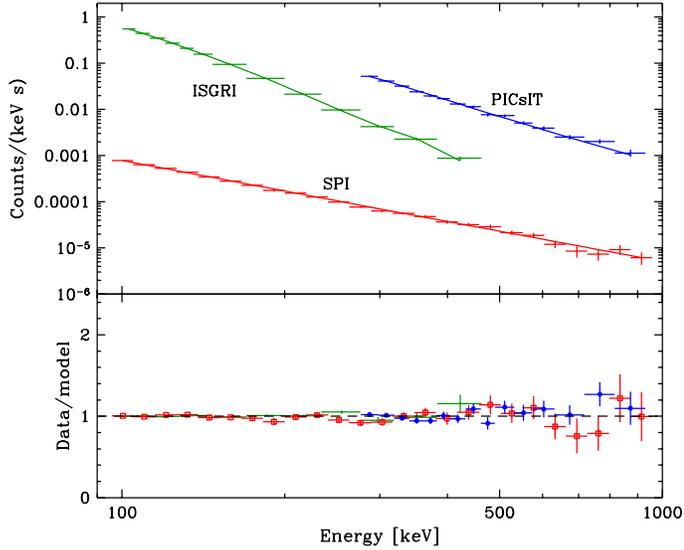}
\caption{Cross-calibration test completed for the INTEGRAL Crab spectra. ISGRI 
and SPI spectra were extracted with the standard OSA 7.0 software, and the PICSIT 
spectrum was extracted with the PPDF method. Results of a power-law fit 
are given in Table \ref{calib}.}
\label{cal1}
\end{figure}

\begin{figure}[!h]
\centering
\includegraphics[width=9cm]{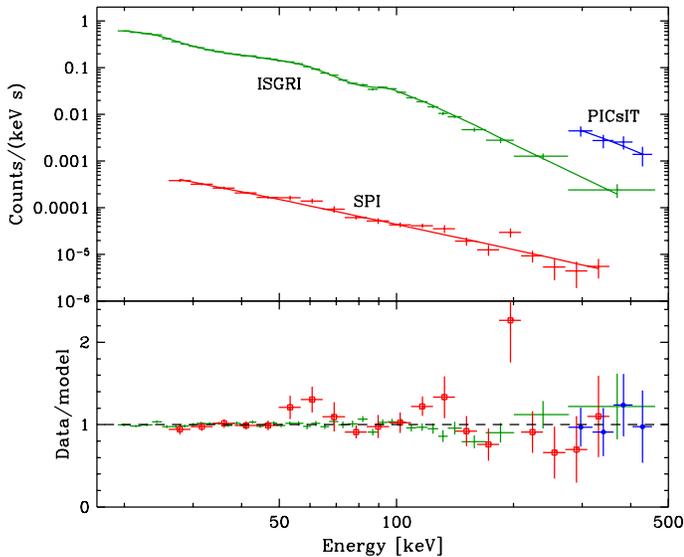}
\caption{Cross-calibration test for the INTEGRAL Cen A spectra. ISGRI and SPI 
spectra were extracted with the standard OSA 7.0 software, and the PICSIT 
spectrum was extracted with the PPDF method. Results of a power-law fit 
are given in Table \ref{calib}.}
\label{cal2}
\end{figure}

Results of spectral fitting performed with the calibration spectra are 
presented in Table \ref{calib} and illustrated in Figs. \ref{cal1} and 
\ref{cal2}. In agreement with the results of tests done with the standard OSA 
software, the PICSIT calibration appears to be reliable, validating the 
instrument model included 
in the Monte Carlo software used for the response simulations. Both the Crab and 
Cen A spectra measured by three INTEGRAL instruments are fully consistent, 
producing similar spectral slopes also when fitted separately. The 
cross-calibration factors measured with respect to SPI obtained for ISGRI and 
PICsIT are both below 1. The PICsIT cross-calibration factor is poorly 
constrained for the Cen A spectra, where the SPI and PICsIT useful bandwidths are 
rather narrow. The relative PICsIT/ISGRI normalization obtained for the Crab of 
0.86 is well above the value of 0.52 quoted for PICsIT spectra extracted with the 
standard OSA software \citep{Foschini07b}. This issue is discussed further in 
Sect. \ref{osa}.

\subsection{Other tests}
\label{other}

\begin{figure}[!h]
\centering
\includegraphics[width=9cm]{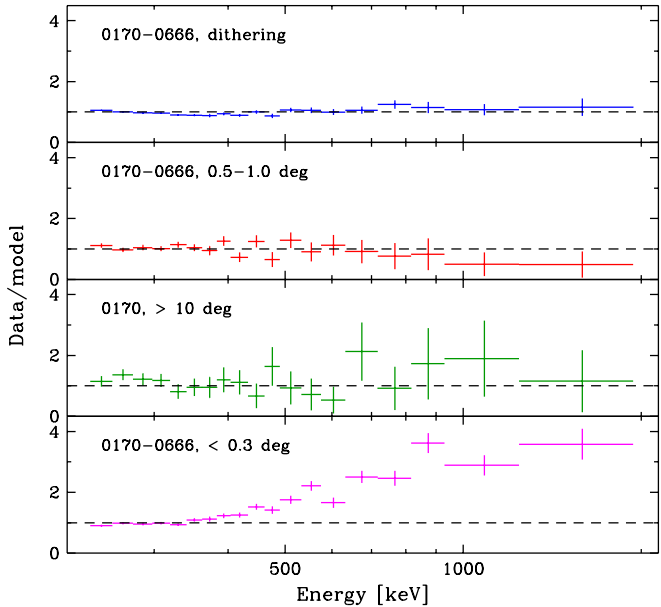}
\caption{Test of the off-axis effect. The plots show the ratio between the data 
and a power-law model for Crab observations from Revs. 0170--0636 with different 
selections of off-axis angles. Model used for each plot was fitted to the 
spectrum made with dithering data. Dithering data comprise all data except 
on-axis data ($<$ 0.3$\degr$). For the $>$ 10$\degr$ selection only the data of 
Rev. 0170 were used.}
\label{offax}
\end{figure}

Besides the verification of the PICsIT calibration, other tests were needed to 
check the performance of the spectral extraction software related to the 
instrument model. One of the main effects to be tested was the dependence of the 
results on the off-axis angle at which the source was observed. Many possible 
problems can arise from different orientations of the spacecraft with respect to 
the object. It is impossible to account for all of them in the instrument model 
included in the spectral extraction or in the response-simulation software. 
Problems with the off-axis observation results were reported for the standard OSA 
software and some corrections to the sky-image deconvolution code were introduced 
starting from the OSA 6.0 release \citep{Foschini07a}. The PIF-based spectral 
extraction presented in this paper accounts for some off-axis effects, as already 
mentioned in Sect. \ref{instrument}. The Crab spectra presented in Fig. 
\ref{offax} show that these corrections are sufficient to ensure stability of the 
results. They were extracted for all data taken during and after Rev. 0170, and 
for several subsets of those data, selected according to the off-axis angle. Both 
large off-axis angle (above 10$\degr$) and small off-axis angle (in the 
0.5$\degr$--1$\degr$ range) spectra are fully consistent with the total spectrum. 
However, it appears that the spectra produced from the on-axis data (off-axis 
angle below $\approx$ 0.3$\degr$) are quite different. They are found to exhibit 
strong spectral hardening observed in this case, starting at an energy of 
$\approx$ 400 keV and resulting in a Crab spectral index $\Gamma$ of around 1.4. 
A similar effect is observed for on-axis spectra extracted with the 
standard OSA 7.0 software. 

Deeper studies of the on-axis phenomenon indicate that its amplitude in 
single-revolution spectra is variable. The hardening is sometimes strong or very 
strong, but sometimes it is quite weak or not present at all. In addition, the 
hardening appears at different energies for individual spectra. A possible 
explanation of at least part of this effect is discussed below, but a full 
explanation will need more tests and additional data. For instance, a 
micro-dithering observation could be useful.  Nevertheless, in practice, sources 
other than the Crab are typically observed with the standard 5$\times$5 or some 
other dithering pattern, and this on-axis effect is not important for them. For 
instance, there are only 19 science windows with Cyg X-1 observed at off-axis 
angles below 0.3$\degr$, among all of the 1726 publicly available pointings for 
that source.

The correctness of the background model provided by the background map is crucial 
for obtaining good PICsIT results. As already mentioned before, such a background 
map cannot be prepared with the staring data because the PIF pattern 
corresponding to the source emission is then included in the background pattern 
and it is not possible to distinguish the signal from the background. 
Tests performed with the maps based on the staring data show the Crab and 
Cen A spectra with a low intensity, similar to the noise spectra derived for 
empty-field observations. This does not mean that the staring data must always be 
neglected. Usually only part of the revolution is consumed by these observations 
and the background map can be prepared with the rest of the data, 
corresponding to the dithering mode. Even when there is no dithering data for a 
given revolution, it is still possible to prepare background maps using 
data from observations carried out some time before or later. In the case of the 
Crab, this solution was tried for Revs. 0039 and 0184. Unfortunately, the change 
in the background pattern after passage through the Earth radiation belts appears 
to be too large and the staring-data spectra obtained for these two observations 
were unsatisfactory. The highest quality spectrum for Rev. 0039 was obtained with 
the background map of Rev. 0044, showing a normal Crab spectral shape but with an 
about 30\% too high normalization. The maps from Revs. 0042, 0043, and 0045 
produced lower quality results, of higher flux and a distorted spectral shape at 
energies higher than $\approx$ 500 keV. This test indicates that the observation 
closest in time was not necessarily the most suitable for preparing a background 
map.

Other Crab staring data were collected together with the dithering data for the
same revolution, allowing us to use the default background map. The spectra 
extracted for both on-axis and off-axis staring data show, in general, the same 
type of behaviour: the flux is too high, and the spectrum is distorted and too 
hard at higher energy. Because the staring observation is sometimes performed in 
different conditions from the dithering ones, e.g., at the beginning or 
end of the revolution, the dithering background map does not always model 
sufficiently well the background for the entire revolution. This can explain 
partly the observed effect, although an incorrect background map normally 
produces a spectrum that is consistent with the noise level. Therefore, the 
higher flux and harder staring spectra, especially those with the source 
observed on-axis, remain unexplained.

\begin{figure}[!h]
\centering
\includegraphics[width=9cm]{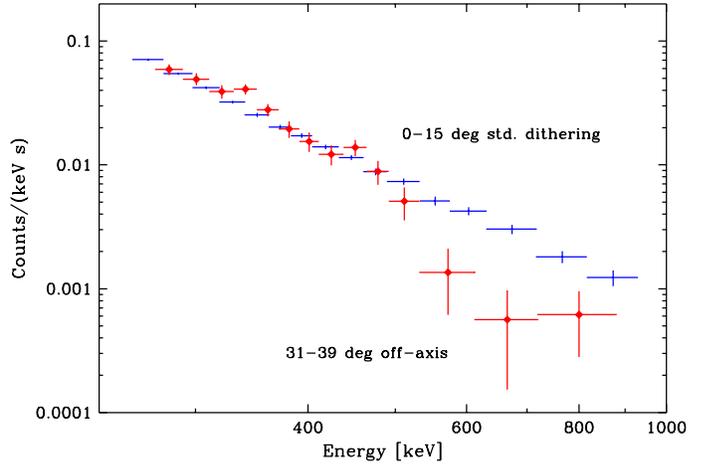}
\caption{Spectra extracted for very large off-axis pointings with the use of the
PIF including  the  IBIS walls and hopper model. Blue crosses show the 
total PICsIT Crab spectrum obtained with the standard PIF for the mask-coded part 
of the shadowgram. Red crosses are the PICsIT Crab spectrum extracted with the 
extended PIF using the data from the Algol observation in Rev. 0220, when the 
Crab was 31--39 degrees off-axis.}
\label{far}
\end{figure}

Figure \ref{far} presents yet another spectral test made for Crab. As mentioned
in Sect. \ref{instrument}, the wall transparency of the IBIS telescope allows a 
quite large fraction of high-energy photons from the source to reach the detector 
when the object is observed at large ($>$ 15$\degr$) off-axis angles. If this 
effect is modelled by the PIF in its extended version, one can in principle 
extract spectra for these observations. However, the quality of the result 
depends on the actual observing strategy because the source-PIF patterns for the 
science windows used to prepare the background map should be sufficiently 
different from each other to allow for disentanglement between the source and 
the background. The far off-axis Crab spectrum shown in Fig. \ref{far} proves
that it is possible to obtain a good result, at least in the energy range up to 
$\approx$500 keV, where no more significantr departure from the normal spectrum 
is seen. This also indicates that the standard response files can be used in such 
a case. Taking into account the fact that for almost all objects observed with 
INTEGRAL, there are more pointings with the off-axis angle in the 
15$\degr$--40$\degr$ range than in the standard dithering $\leq$ 15$\degr$ range, 
plenty additional data can be used. This technique needs further studies to be 
carried out together with the investigation aimed at finding the most reliable 
method of verifying the applicability of the background map, mentioned at the end 
of Sect. \ref{maps}. 

\subsection{Comparison with the OSA 7.0 results}
\label{osa}

The spectra extracted using the PPDF method have to be compared with the spectra 
extracted with the standard PICsIT analysis software distributed within the OSA 
package. Because the PIF-based spectral extraction software included in OSA 7.0 
is still unreliable \citep{Foschini07b}, OSA spectra were prepared from the sky 
mosaic images. Both methods of creating mosaic images were tested: the first  
applies the OSA tool {\sl ip\_skymosaic}, the second utilizes the {\sl varmosaic} 
tool distributed within the HEASOFT package. The data analysis performed  
using the OSA followed exactly the instructions given in the document ``PICsIT 
analysis made easy" prepared by the IBIS Team \citep{Foschini07c}. Mosaic images 
based on the single-event data were created in the 8 default energy bins, 
corresponding to the channels of the rebinned response matrix {\sl 
pics\_srmf\_grp\_0005.fits}. All parameters were set to the default values, 
except for {\sl PICSIT\_inCorVar}, which was set to 1 to take into account an 
uncertainty in the background map. Crab spectra were extracted for the same sets 
of science windows as for the PPDF method, in order to test the software 
performance in various conditions. 

\begin{figure}[!h]
\centering
\includegraphics[width=9cm]{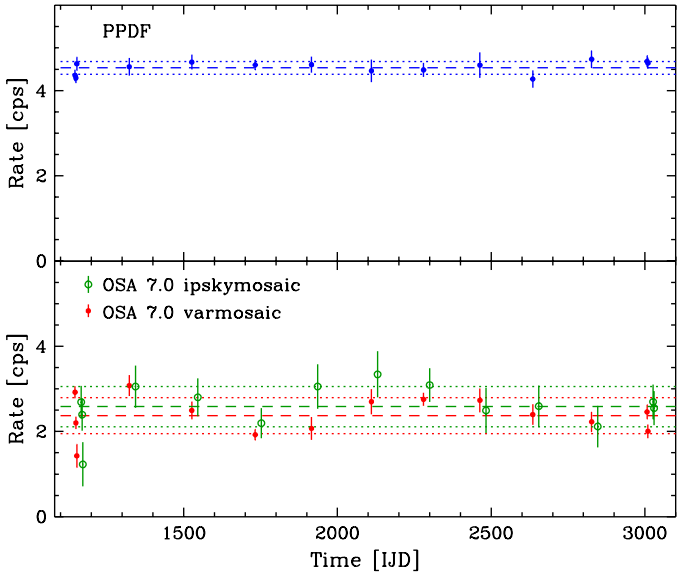}
\caption{Comparison between the Crab results obtained with the PPDF method and
with the standard OSA 7.0 software. The upper part shows the Crab light curve in
the 255--340 keV band extracted with the PPDF method. In the lower part, Crab
light curves in the 252--336 keV band made with OSA 7.0 are shown. OSA mosaic 
images were obtained with two alternative methods, one is the standard OSA tool 
{\sl ip\_sky\_mosaic} and the second is the {\sl varmosaic} tool distributed in 
the HEASOFT package. Dashed lines show the mean count rate level and dotted lines 
one standard deviation limits. Points showing the {\sl ip\_sky\_mosaic} results 
are slightly shifted to the right for a better visibility.}
\label{lcr}
\end{figure}

Figure \ref{lcr} and Table \ref{light} present a comparison of the OSA and PPDF 
Crab count rates in a broad, low-energy band. For OSA this is the second default 
channel between 252 and 336 keV, i.e. the lowest energy channel unaffected by the 
track events. For the PPDF 22-bin spectra, channels 2-5 were merged, giving the 
count rate in the 255--340 keV band. The Crab light curve extracted with the PPDF 
method is remarkably stable and the results are much more precise than those 
obtained with the OSA software. The Crab count rates computed by the PPDF method 
are almost two times higher than those read from the OSA sky mosaic images. 
Since there are so many different elements in these two approaches, it is 
difficult to identify a main reason for such a large count-rate deviation. 
Possible explanation can be: a difference between the PIF model and IBIS mask 
model used in the image deconvolution, different background maps, and 
different approach used to calculate the net source-count rate. Moreover, the 
fact that the count rates extracted from mosaic images created using {\sl 
ip\_skymosaic} and {\sl varmosaic} sometimes differ, shows that the mosaic 
results can be unstable. Although the {\sl PICSIT\_inCorVar} parameter was set to 
1 in the sky image deconvolution, the uncertainty in the count rates extracted 
from the mosaic produced with the {\sl varmosaic} tool are clearly too small 
compared to the scatter of results.

A detailed analysis of the standard OSA 7.0 PICsIT spectra is beyond the scope 
of this paper. Nevertheless, we completed spectral fitting for about 20 different 
Crab spectra extracted for different observing conditions, which were fitted 
separately first and then together, for both mosaicing tools. In addition, the 
long-exposure spectra from mosaics containing data from many revolutions were 
prepared to be compared with PPDF spectra extracted for the same data 
set. The default energy bins defined in the OSA software are too wide to allow 
a detailed spectral analysis. Therefore, only some basic tests were completed
by fitting a power-law model to all 8 bins (203--6720 keV) or to the limited band 
with channels 1 and 8 excluded. Spectra of shorter exposure times, e.g. those 
corresponding to a single revolution, were fitted by retaining the relative 
normalization as a free parameter.

\begin{table}
\begin{center}
\caption{Parameters of the Crab light curves obtained with the PPDF and OSA 7.0
(I - {\sl ip\_skymosaic}, V - {\sl varmosaic}) methods.} 
\label{light}
\begin{tabular}{ccccc}
\hline\hline
Method & Mean & Std. dev. & Rel. error & $\chi^{2}$/NDF \\
\hline
 PPDF   & 4.535 & 0.155 &  3.4\% & 12.0/13 \\
 OSA, I & 2.372 & 0.422 & 17.8\% & 65.5/13 \\
 OSA, V & 2.585 & 0.473 & 18.3\% & 15.0/13 \\
\hline\hline 
\end{tabular}  
\end{center}
\end{table}

In general, the Crab spectral-fitting results were consistent for all three 
methods, i.e. the PPDF and both OSA-based mosaics. The relative normalization of 
the short-exposure Crab spectra was similar for each method, confirming, for 
example, the effect of higher count rates, as measured for on-axis spectra. 
However, these numbers and the spectral indexes are affected by significantly 
more scatter for the OSA results than for the PPDF results. The most extreme 
cases are staring spectra from Rev. 0102, which have a $\Gamma$ value of 3.93, 
and Rev. 0045 spectra with a $\Gamma$ of 1.43. Both observations were performed 
with a relatively small Sun off-axis angle and may have been affected by a higher 
and more variable background than usual. For other spectra, $\Gamma$ values are 
within the range 1.5--3.1 and the relative normalization parameters are within 
the range 0.5--1.4. The results of the spectral fitting for all dithering spectra 
together are presented in Table \ref{osafit}. All spectra fitted in the 300--1000 
keV band had a slightly smaller spectral index than obtained for the SPI spectra. 
This suggests that the steepness of the PICsIT ARF was overestimated in this 
energy range. Fluxes obtained with the PPDF method are about a factor of two 
higher than those of the OSA spectra, in agreement with the result for a narrower 
band, presented in Table \ref{light}. 

\begin{table}
\begin{center}
\caption{Parameters of the Crab spectral fitting obtained for dithering spectra
extracted with the PPDF and OSA 7.0 (I - {\sl ip\_skymosaic}, V - {\sl 
varmosaic}) methods. The PPDF spectrum was fitted in the 298--1938 keV range, OSA 
spectra in the 252--1848 keV range. Flux in the 300--1000 keV band is given in 
ergs cm$^{-2}$ s$^{-1}$. Rows 2 and 3 show the OSA results obtained when single
revolution spectra were fitted together with a free relative normalization, rows 
4 and 5 correspond to the fit made for spectra extracted from all-data mosaics.} 
\label{osafit}
\begin{tabular}{ccccc}
\hline\hline
Method & $\Gamma$ & Norm. & Flux & Red. $\chi^{2}$ \\
\hline
 PPDF   & 2.07$\pm$0.05 & 6.2$\pm$1.7 & $7.8\times 10^{-9}$ & 0.80 \\
 OSA, I & 2.02$\pm$0.16 & 2.7$\pm$2.5 & $4.6\times 10^{-9}$ & 0.19 \\
 OSA, V & 2.05$\pm$0.07 & 3.4$\pm$1.3 & $4.1\times 10^{-9}$ & 0.97 \\
 OSA, I & 1.98$\pm$0.21 & 1.6$\pm$1.8 & $3.3\times 10^{-9}$ & 0.14 \\
 OSA, V & 2.02$\pm$0.07 & 2.3$\pm$0.9 & $4.0\times 10^{-9}$ & 2.39 \\
\hline\hline 
\end{tabular}  
\end{center}
\end{table}

\begin{figure}[!h]
\centering
\includegraphics[width=9cm]{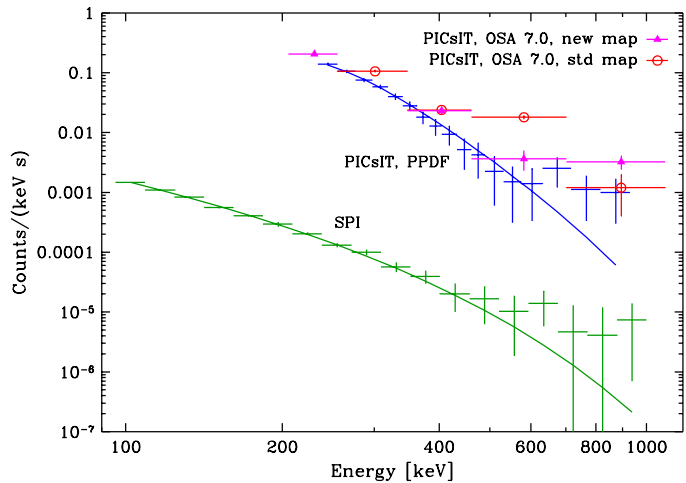}
\caption{Comparison between the SPI and PICsIT spectra of Cyg X-1 during the
September 2006 outburst. The model was fitted to the SPI and PICsIT PPDF spectra 
in the 100-600 keV band.}
\label{flare}
\end{figure}

\begin{figure}[!h]
\centering
\includegraphics[width=9cm]{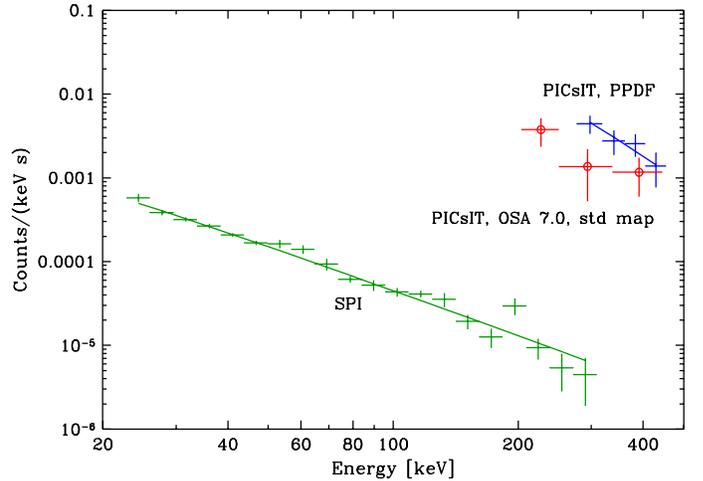}
\caption{Comparison between the SPI and PICsIT spectra of Cen A observations in
2003--2004. The model was fitted to the SPI and PICsIT PPDF spectra.}
\label{cena}
\end{figure}

PPDF and OSA spectra were also compared for two other sources, and the results 
are shown in Figs. \ref{flare} and \ref{cena}. The \object{Cyg X-1} observation 
completed in September 2006 (Rev. 0482) was special because the source was then 
brighter than the Crab. Unfortunately, this was a routine Galactic Plane Scan 
observation and the object was always at least 9$\degr$ off-axis. Therefore, the 
results were not optimal for studies of the physics of this Cyg X-1 brightening 
but, on the other hand, provided a good opportunity to test the OSA performance 
under special conditions. The model plotted in Fig. \ref{flare} is a cut-off 
power-law fitted only to the SPI and PICsIT PPDF spectra, in the 100--600 keV 
band. Because this observation took place well after the period covered by the 
data used to prepare the default OSA PICsIT background maps, a revised map was 
prepared for the default OSA energy bins with the data taken from 
revolution 0482. However, both types of PICsIT OSA spectra extracted for this 
observation, i.e. using either the default or new map, were of quite low quality. 
There was no detection (negative count rate) in the first and second band of the 
default-map and new-map spectra, respectively. The default-map spectrum exhibited 
large deviations from the model, and both of the OSA spectra exhibit high 
levels of flux, above that of the PPDF spectrum, whereas the Crab results showed 
the opposite trends. Noise-level studies for an observation of similar exposure 
time (see Sect. \ref{limits}) indicated reliable source detection for energies of 
up to 600 keV. Thus, the spectral hardening observed in the PPDF spectrum above 
that energy was only the noise fluctuation. A detailed analysis and discussion of 
the INTEGRAL results for the Cyg X-1 flare observed in September 2006 were 
presented in \citet{Malzac08}.

The third object, for which the OSA PICsIT spectra were tested was Cen A, to 
check whether is was possible to study a source much weaker than Crab or Cyg X-1. 
The same data as in the case of the calibration studies were used, and the OSA 
mosaic was made with the {\sl varmosaic} tool only. The first two channels of the 
OSA PICsIT spectrum are in agreement with the model, assuming a relative 
normalization similar to the Crab case. Nevertheless, the OSA spectrum is much 
less precise than that extracted with the PPDF method and provides no additional 
information to that derived using the SPI spectrum.

\subsection{PICsIT detections}

We review PICsIT spectra extracted for objects of various
types. Although most of these results are only preliminary and were obtained
without a careful analysis and data selection, they illustrate the type of 
information that PICsIT can provide. The number of objects with relatively
strong emission within the PICsIT energy range between several hundreds of keV and
several MeV is very small. The dominant class of sources are Galactic binary systems with
a black hole candidate. Three other sources detected until now were the Crab and 
\object{PSR B1509-58} pulsars and a single AGN, Cen A. Several objects were 
detected only during outbursts of several days or weeks because their persistent 
emission was too weak for PICsIT. Currently, there are 12 firmly detected sources 
(with $>$ 6-sigma detections in the 277--361 keV band) among about 30 tested.  

\begin{figure}[!h]
\centering
\includegraphics[width=9cm]{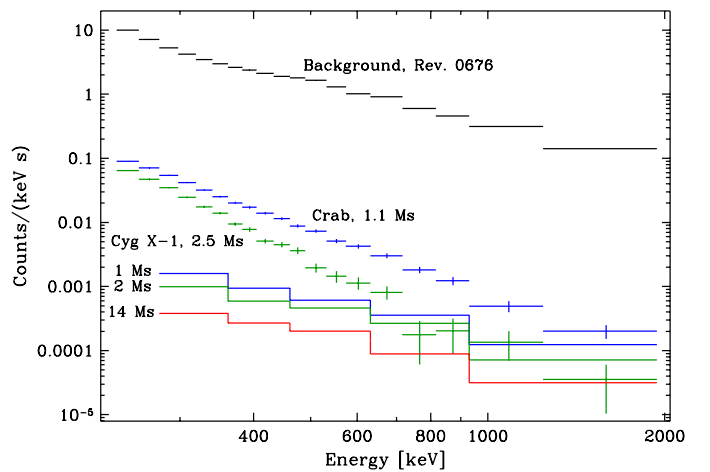}
\caption{PICsIT spectra of the two brightest objects, Crab and Cyg X-1, compared 
to the PICsIT background spectrum and 3-sigma upper noise limits for 1, 2 and
14 Ms of effective exposure time.}
\label{strong}
\end{figure}

Figure \ref{strong} illustrates the scale of the problem for a detection of 
emission in the soft $\gamma$-ray domain. Cosmic-ray induced internal background 
of the detector is usually orders of magnitude higher than the radiation 
of the brightest sources. In the case of PICsIT, the background is a factor of 
100 stronger than the Crab emission around 300 keV, and 1000 times stronger at 
several MeV. Due to the coded-mask technique and dithering observation strategy, 
it remains possible to model the background to a precision allowing objects many 
times weaker than Crab to be detected, provided that there is sufficient exposure 
time. Crab was observed only during one or two revolutions twice a year, and the 
total exposure time of about 1 Ms provides a clear detection up to 2 MeV. Cyg X-1 
was observed far more frequently but its steeper spectrum allowed us to detect 
the source only up to about 700 keV. Because more data will become public from 
the INTEGRAL Key Programme observations of the Cygnus region, future analysis 
should yield a spectrum reaching at least 1 MeV, according to the noise limits 
shown in Fig. \ref{strong}.

\begin{figure}[!h]
\centering
\includegraphics[width=9cm]{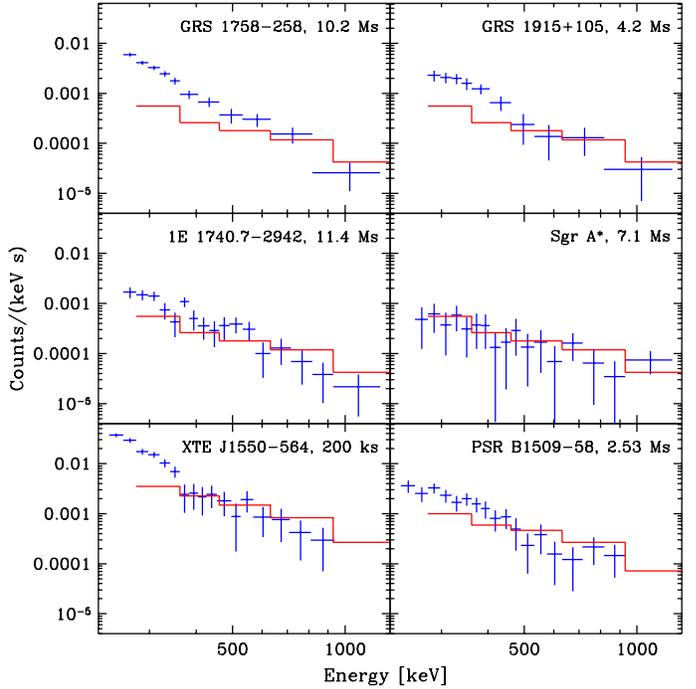}
\caption{ Examples of PICsIT spectra extracted for several Galactic objects. 
Spectrum extracted for the position of Sgr A* presents situation, when there is
no detection. Spectra in four upper panels are compared with the 3-sigma 
upper noise limits derived for an effective exposure time of 10 Ms. Spectrum of 
XTE J1550-564 measured during outburst is compared with the 3-sigma upper noise 
limits computed for an effective exposure time of 200 ks and the PSR B1509-58 
spectrum with the 3-sigma upper noise limits for an effective exposure time of 
2 Ms.}
\label{objects}
\end{figure}

The spectra of three other persistent sources that can be well studied 
with PICsIT are presented in Fig. \ref{objects}. Due to a long exposure time, the 
results obtained for the microquasar \object{GRS 1758-258} were of overhelmingly 
higher quality than any obtained for this object with any other, past, or 
contemporaneous, instrument. As data continues to arrive in the next few years 
for this Galactic Bulge source it will be possible to study the emission 
properties up to about 600 keV, with the possibility of subdividing the data 
according to the object state. The other bright microquasar, 
\object{GRS 1915+105}, would need a more sophisticated analysis, but preliminary 
results indicate rather unambiguously a harder spectrum in the soft $\gamma$-ray 
band than for GRS 1758-258, and a cut-off at around 500 keV. The third object 
with the spectrum shown in Fig. \ref{objects}, \object{1E 1740.7-2942}, exhibits 
weaker emission in the PICsIT energy band, and more data and a more careful 
analysis will be needed to obtain a high-quality spectrum. Nevertheless, the 
final spectrum should not differ significantly from the current one, showing yet 
another emission pattern with a hard spectral slope. The other spectrum of Fig. 
\ref{objects} was extracted for the position of the Galactic Centre radio-source 
\object{Sgr A*}. Taking into account that no strong hard X-ray emission 
originates in this point (ISGRI observations of \object{IGR J17456-2901} reveal 
a 7 mCrab source in the 60-150 keV band \citep{Kuulkers07}), it is unsurprising 
that the PICsIT spectrum is fully consistent with the noise spectrum. The 
expected exposure time of in excess of 20 Ms at the end of the INTEGRAL mission 
will allow us to derive the most reliable ever upper limit to the emission from 
the centre of our Galaxy. We emphasize that for both 1E 1740.7+2942 and Sgr A* 
located in a region of strong diffuse emission observed in the 200-500 keV range 
by SPI \citep{Bouchet08}, PICsIT, due to its high angular resolution, is the only 
instrument capable of resolving point sources without the ambiguity affecting SPI 
and Suzaku/HXD results.

Besides the bright persistent soft $\gamma$-ray emitters mentioned above, several 
objects are too weak to be detected by PICsIT in the normal state but bright 
enough during relatively long periods of outbursts. Five sources of this type 
have been detected with PICsIT. Figure \ref{objects} presents the spectrum of one 
of these sources, XTE J1550-564. The other four are black hole 
candidates: \object{GRO J1655-40}, \object{GX 339-4}, \object{IGR J17497-2821}, 
and \object{SWIFT J1753.5-0127}. The last spectrum shown in Fig. \ref{objects} 
illustrates detection of emission from the PSR B1509-58 pulsar system. This 
source was detected by PICsIT with the standard OSA software \citep{Kuiper05}, 
although no details were then given.
 
\subsection{Future prospects}

The spectral extraction method presented in this paper provides valuable and 
reliable results, as shown in the previous subsections. Notwithstanding, possible
ways of improving the PICsIT spectra remain, which would increase the sensitivity 
of the instrument and the stability of the results. The first possibility is 
application of a spectral extraction also to the multiple  events. Although 
several times fewer photons are registered in multiple than single events 
below 1 MeV, the instrumental background for multiple events is also several 
times lower. This can allow the sensitivity to be improved by about 20\% above 
300 keV. More careful modelling of the background and more careful selection of 
the data will produce more stable source count-rates, smoother spectra, and more 
precise light curves. This goal can be achieved by developing tools to verify the 
adequacy and stability of the background during a single revolution or a sequence 
of observations. Currently, a simple way of achieving a more accurate background 
model is to prepare two or three background maps for a given revolution.

Further improvements in sensitivity, of up to a factor 2, can be achieved with 
the development of an efficient tool to verify the background-map applicability 
to the data of other revolutions. This would allow us to extract the results for 
sources observed in staring mode or at very large off-axis angles, when the 
standard map produced for a given revolution does not allow us to distinguish 
between the source and the background. In this way, the total exposure time for 
many objects could be increased by a factor of 2--4, which would be important for 
extragalatic objects observed rarely with INTEGRAL. 

\section{Summary and conclusions}
\label{summary}

We have developed a comprehensive method for extracting PICsIT spectra from  
spectral-imaging single-event data, which enables us to derive information to 
the physical limitations of the instrument. This was achieved mainly by 
application of a correct description of the Poisson-distributed data and careful 
modelling of the detector background. 

We adopted a technique that handles the Poisson probability density functions in 
a Bayesian manner for the first time in extracting the count rates from a 
$\gamma$-ray detector.  The effectiveness of this technique in the `low number 
of counts' regime and its superiority over other methods was proven with 
exhaustive tests.

In our spectral extraction technique we employed a background model that was as
close as possible to the data because it consisted of a map of the mean count rate 
measured for each pixel during a given object observation. This approach is 
justified for a coded-mask instrument operating in a dithering mode, especially 
when the background is orders of magnitude stronger than the source emission. 

The detection reliability was verified by detailed studies of fake source spectra 
extracted from empty field observations. This approach provided the most reliable 
estimate of the noise level possible, which accounted for true background 
fluctuations and other systematic effects related to the instrument and spectral 
extraction method. We recommend similar tests for other instruments, because the 
standard analytical calculation of the detection limit can often underestimate 
the true noise limit.

We tested the performance of the new PICsIT spectral extraction method by 
comparing our results with those from other INTEGRAL instruments. Using the 
standard PICsIT response files, we found good agreement between the spectral
fitting results for data from PICsIT and both SPI and ISGRI, obtained for the 
Crab, Cen A, and Cyg X-1. PICsIT appears to be well calibrated, although some 
tuning of the response files may be needed below 300 keV and above 1 MeV.

Confronted with the standard OSA software results, count rates computed with the 
new method are far more stable and about a factor of two higher, in close 
agreement with results derived for data from other INTEGRAL instruments. Crab 
spectra extracted with both OSA and the new method are described well by a 
model of the same spectral slope. However, for weaker objects and for 
observations made at large off-axis angles, we were unable to obtain high quality
spectra with the standard software.

The new spectral extraction method was applied to date for about 30 objects 
observed with INTEGRAL. Eight new sources were added to the four already 
reported, which were detected with the standard PICsIT software. A refined 
analysis and an expanding data set should provide at least several detections 
more. Due to a very long exposure time, the spectrum obtained for the microquasar 
GRS 1758-258 reached an unprecedented quality and could be extended to higher 
energies up to about 700 keV. Preliminary spectra extracted for several other 
sources, in particular those observed during outbursts, also provide information 
unachievable for any other instrument currently operating.

Despite its unique advantages, PICsIT has a limited sensitivity when compared to 
the OSSE or BATSE detectors. The main reason for this is the smaller thickness 
of the detector layer, 3 cm for PICsIT, compared to 17.8 cm for OSSE and 7.6 cm 
for BATSE. In addition, the PICsIT ability to detect photons is decreased by the 
pixellated structure of the detector and the logic discriminating single and 
multiple events. Nevertheless, it seems that a coded-mask instrument is the most
appropriate solution for soft $\gamma$-ray astronomy, because this technique 
offers at the same time a large field of view, good angular resolution and direct 
measurement of the background when dithering observations are used.

\begin{acknowledgements}
This work was supported by the Polish MNiSW grants 1P03D01128 and NN203065933. 
The author would like to thank Marc T\"urler for reading the manuscript and 
valuable comments. The comments by the anonymous referee improved the
manuscript substantially.
\end{acknowledgements}

\bibliographystyle{aa}
\bibliography{pl}

\begin{thebibliography}{50}
\expandafter\ifx\csname natexlab\endcsname\relax\def\natexlab#1{#1}\fi

\bibitem[{{Arnaud}(1996)}]{Arnaud96}
{Arnaud}, K.~A. 1996, in Astronomical Society of the Pacific Conference Series,
  Vol. 101, Astronomical Data Analysis Software and Systems V, ed. G.~H.
  {Jacoby} \& J.~{Barnes}, 17

\bibitem[{{Atti{\'e}} {et~al.}(2003){Atti{\'e}}, {Cordier}, {Gros}, {Laurent},
  {Schanne}, {Tauzin}, {von Ballmoos}, {Bouchet}, {Jean}, {Kn{\"o}dlseder},
  {Mandrou}, {Paul}, {Roques}, {Skinner}, {Vedrenne}, {Georgii}, {von Kienlin},
  {Lichti}, {Sch{\"o}nfelder}, {Strong}, {Wunderer}, {Shrader}, {Sturner},
  {Teegarden}, {Weidenspointner}, {Kiener}, {Porquet}, {Tatischeff}, {Crespin},
  {Joly}, {Andr{\'e}}, {Sanchez}, \& {Leleux}}]{Attie2003}
{Atti{\'e}}, D., {Cordier}, B., {Gros}, M., {et~al.} 2003, \aap, 411, L71

\bibitem[{B\'elanger(2008)}]{Belanger08}
B\'elanger, G. 2008, {I}BIS Observer's Manual,
  http://integral.esac.esa.int/AO6/ AO6\_IBIS\_om.pdf

\bibitem[{{Bird} {et~al.}(2003){Bird}, {Bazzano}, {Ferguson}, {La Rosa},
  {Malaguti}, \& {Ubertini}}]{Bird03}
{Bird}, A.~J., {Bazzano}, A., {Ferguson}, C., {et~al.} 2003, \aap, 411, L197

\bibitem[{{Bouchet} {et~al.}(2008){Bouchet}, {Jourdain}, {Roques}, {Strong},
  {Diehl}, {Lebrun}, \& {Terrier}}]{Bouchet08}
{Bouchet}, L., {Jourdain}, E., {Roques}, J.-P., {et~al.} 2008, \apj, 679, 1315

\bibitem[{{Cadolle Bel} {et~al.}(2006){Cadolle Bel}, {Sizun}, {Goldwurm},
  {Rodriguez}, {Laurent}, {Zdziarski}, {Foschini}, {Goldoni}, {Gouiff{\`e}s},
  {Malzac}, {Jourdain}, \& {Roques}}]{Cadolle06}
{Cadolle Bel}, M., {Sizun}, P., {Goldwurm}, A., {et~al.} 2006, \aap, 446, 591

\bibitem[{{Cash}(1979)}]{Cash79}
{Cash}, W. 1979, \apj, 228, 939

\bibitem[{{Churazov} {et~al.}(2007){Churazov}, {Sunyaev}, {Revnivtsev},
  {Sazonov}, {Molkov}, {Grebenev}, {Winkler}, {Parmar}, {Bazzano}, {Falanga},
  {Gros}, {Lebrun}, {Natalucci}, {Ubertini}, {Roques}, {Bouchet}, {Jourdain},
  {Kn{\"o}dlseder}, {Diehl}, {Budtz-Jorgensen}, {Brandt}, {Lund},
  {Westergaard}, {Neronov}, {T{\"u}rler}, {Chernyakova}, {Walter}, {Produit},
  {Mowlavi}, {Mas-Hesse}, {Domingo}, {Gehrels}, {Kuulkers}, {Kretschmar}, \&
  {Schmidt}}]{Churazov07}
{Churazov}, E., {Sunyaev}, R., {Revnivtsev}, M., {et~al.} 2007, \aap, 467, 529

\bibitem[{{Courvoisier} {et~al.}(2003){Courvoisier}, {Walter}, {Beckmann},
  {Dean}, {Dubath}, {Hudec}, {Kretschmar}, {Mereghetti}, {Montmerle},
  {Mowlavi}, {Paltani}, {Preite Martinez}, {Produit}, {Staubert}, {Strong},
  {Swings}, {Westergaard}, {White}, {Winkler}, \& {Zdziarski}}]{Courvoisier03}
{Courvoisier}, T.~J.-L., {Walter}, R., {Beckmann}, V., {et~al.} 2003, \aap,
  411, L53

\bibitem[{{Di Cocco} {et~al.}(2003){Di Cocco}, {Caroli}, {Celesti}, {Foschini},
  {Gianotti}, {Labanti}, {Malaguti}, {Mauri}, {Rossi}, {Schiavone},
  {Spizzichino}, {Stephen}, {Traci}, \& {Trifoglio}}]{DiCocco03}
{Di Cocco}, G., {Caroli}, E., {Celesti}, E., {et~al.} 2003, \aap, 411, L189

\bibitem[{{Foschini}(2004)}]{Foschini04}
{Foschini}, L. 2004, {O}SA 4.0 Improvements for PICsIT, ISDC Newsletter No. 15,
  http://isdc.unige.ch/Newsletter/N15

\bibitem[{{Foschini}(2005)}]{Foschini05}
{Foschini}, L. 2005, {T}he outburst of XTE J1550-564 in 2003 as seen by IBIS,
  ISDC Newsletter No. 17, http://isdc.unige.ch/Newsletter/N17

\bibitem[{{Foschini}(2007{\natexlab{a}})}]{Foschini07c}
{Foschini}, L. 2007{\natexlab{a}}, {P}ICSIT Data Analysis made easy,
  http://www.iasf-bologna.inaf.it/~foschini/OSAP/picsit\_data\_analysis.html

\bibitem[{{Foschini}(2007{\natexlab{b}})}]{Foschini07a}
{Foschini}, L. 2007{\natexlab{b}}, {I}BIS/PICsIT Novelties in OSA 6.0, ISDC
  Newsletter No. 20, http://isdc.unige.ch/Newsletter/N20

\bibitem[{{Foschini} {et~al.}(2007){Foschini}, {Bianchin}, {Goldwurm}, {Gros},
  {Malaguti}, {Laurent}, \& {Natalucci}}]{Foschini07b}
{Foschini}, L., {Bianchin}, V., {Goldwurm}, A., {et~al.} 2007, {I}BIS/PICsIT
  Instrument Specific Software, Scientific Validation Report,
  http://isdc.unige.ch/Soft
  /download/osa/osa\_doc/prod/osa\_sci\_val\_picsit-6.0.pdf

\bibitem[{{Frontera} {et~al.}(1997){Frontera}, {Costa}, {dal Fiume}, {Feroci},
  {Nicastro}, {Orlandini}, {Palazzi}, \& {Zavattini}}]{Frontera97}
{Frontera}, F., {Costa}, E., {dal Fiume}, D., {et~al.} 1997, \aaps, 122, 357

\bibitem[{{Gehrels} {et~al.}(2004){Gehrels}, {Chincarini}, {Giommi}, {Mason},
  {Nousek}, {Wells}, {White}, {Barthelmy}, {Burrows}, {Cominsky}, {Hurley},
  {Marshall}, {M{\'e}sz{\'a}ros}, {Roming}, {Angelini}, {Barbier}, {Belloni},
  {Campana}, {Caraveo}, {Chester}, {Citterio}, {Cline}, {Cropper}, {Cummings},
  {Dean}, {Feigelson}, {Fenimore}, {Frail}, {Fruchter}, {Garmire}, {Gendreau},
  {Ghisellini}, {Greiner}, {Hill}, {Hunsberger}, {Krimm}, {Kulkarni}, {Kumar},
  {Lebrun}, {Lloyd-Ronning}, {Markwardt}, {Mattson}, {Mushotzky}, {Norris},
  {Osborne}, {Paczynski}, {Palmer}, {Park}, {Parsons}, {Paul}, {Rees},
  {Reynolds}, {Rhoads}, {Sasseen}, {Schaefer}, {Short}, {Smale}, {Smith},
  {Stella}, {Tagliaferri}, {Takahashi}, {Tashiro}, {Townsley}, {Tueller},
  {Turner}, {Vietri}, {Voges}, {Ward}, {Willingale}, {Zerbi}, \&
  {Zhang}}]{Gehrels04}
{Gehrels}, N., {Chincarini}, G., {Giommi}, P., {et~al.} 2004, \apj, 611, 1005

\bibitem[{{Goldwurm} {et~al.}(2003){Goldwurm}, {David}, {Foschini}, {Gros},
  {Laurent}, {Sauvageon}, {Bird}, {Lerusse}, \& {Produit}}]{Goldwurm03}
{Goldwurm}, A., {David}, P., {Foschini}, L., {et~al.} 2003, \aap, 411, L223

\bibitem[{{Hubbell} \& {Seltzer}(1996)}]{Hubbell96}
{Hubbell}, J.~H. \& {Seltzer}, S.~M. 1996, {T}ables of X-ray Mass Attenuation
  coefficients, avalaible at http://physics.nist.gov/
  PhysRefData/XrayMassCoef/cover.html

\bibitem[{{Johnson} {et~al.}(1993){Johnson}, {Kurfess}, {Purcell}, {Matz},
  {Ulmer}, {Strickman}, {Murphy}, {Grabelsky}, {Kinzer}, {Share}, {Cameron},
  {Kroeger}, {Maisack}, {Jung}, {Jensen}, {Clayton}, {Leising}, {Grove}, \&
  {Dyer}}]{Johnson93}
{Johnson}, W.~N., {Kurfess}, J.~D., {Purcell}, W.~R., {et~al.} 1993, \aaps, 97,
  21

\bibitem[{{Jourdain} {et~al.}(2008){Jourdain}, {Gotz}, {Westergaard},
  {Natalucci}, \& {Roques}}]{Jourdain08}
{Jourdain}, E., {Gotz}, D., {Westergaard}, N., {Natalucci}, L., \& {Roques}, J.
  2008, arXiv:0810.0646v1

\bibitem[{{Kuiper}(2005)}]{Kuiper05}
{Kuiper}, L. 2005, in the IBIS/PICsIT Source Catalog web page,
  http://www.iasfbo.inaf.it/extras/Research/INTEGRAL/Catalogue/picsit
  \_soucat.html

\bibitem[{Kuulkers(2005)}]{Kuulkers05}
Kuulkers, E. 2005, http://integral.esa.int/newsletters/ISOC\_newsletter\_15.pdf

\bibitem[{Kuulkers(2006)}]{Kuulkers06}
Kuulkers, E. 2006, {I}BIS Observer's Manual, avalaible at the ESA INTEGRAL
  documentation web page, http://integral.esa.int/AO4/documentation.html

\bibitem[{{Kuulkers} {et~al.}(2007){Kuulkers}, {Shaw}, {Paizis}, {Chenevez},
  {Brandt}, {Courvoisier}, {Domingo}, {Ebisawa}, {Kretschmar}, {Markwardt},
  {Mowlavi}, {Oosterbroek}, {Orr}, {R{\'{\i}}squez}, {Sanchez-Fernandez}, \&
  {Wijnands}}]{Kuulkers07}
{Kuulkers}, E., {Shaw}, S.~E., {Paizis}, A., {et~al.} 2007, \aap, 466, 595

\bibitem[{{Labanti} {et~al.}(2003){Labanti}, {Di Cocco}, {Ferro}, {Gianotti},
  {Mauri}, {Rossi}, {Stephen}, {Traci}, \& {Trifoglio}}]{Labanti03}
{Labanti}, C., {Di Cocco}, G., {Ferro}, G., {et~al.} 2003, \aap, 411, L149

\bibitem[{{Lampton} {et~al.}(1976){Lampton}, {Margon}, \& {Bowyer}}]{Lampton76}
{Lampton}, M., {Margon}, B., \& {Bowyer}, S. 1976, \apj, 208, 177

\bibitem[{{Lebrun} {et~al.}(2003){Lebrun}, {Leray}, {Lavocat}, {Cr{\'e}tolle},
  {Arqu{\`e}s}, {Blondel}, {Bonnin}, {Bou{\`e}re}, {Cara}, {Chaleil}, {Daly},
  {Desages}, {Dzitko}, {Horeau}, {Laurent}, {Limousin}, {Mathy}, {Mauguen},
  {Meignier}, {Molini{\'e}}, {Poindron}, {Rouger}, {Sauvageon}, \&
  {Tourrette}}]{Lebrun03}
{Lebrun}, F., {Leray}, J.~P., {Lavocat}, P., {et~al.} 2003, \aap, 411, L141

\bibitem[{{Loredo}(1990)}]{Loredo90}
{Loredo}, T.~J. 1990, in Maximum Entropy and Bayesian Methods, ed. P.~F.
  {Fougere}, 81

\bibitem[{{Lubi{\'n}ski}(2004)}]{Lubinski04}
{Lubi{\'n}ski}, P. 2004, \mnras, 350, 596

\bibitem[{Lubi\'nski(2007)}]{Lubinski07}
Lubi\'nski, P. 2007, {T}he new IBIS off-axis correction in OSA 6.0, ISDC
  Newsletter No. 20, http://isdc.unige.ch/Newsletter/N20

\bibitem[{{Makishima} {et~al.}(2008){Makishima}, {Takahashi}, {Yamada}, {Done},
  {Kubota}, {Dotani}, {Ebisawa}, {Itoh}, {Kitamoto}, {Negoro}, {Ueda}, \&
  {Yamaoka}}]{Makishima08}
{Makishima}, K., {Takahashi}, H., {Yamada}, S., {et~al.} 2008, \pasj, 60, 585

\bibitem[{{Malaguti} {et~al.}(2003{\natexlab{a}}){Malaguti}, {Bazzano},
  {Beckmann}, {Bird}, {Del Santo}, {Di Cocco}, {Foschini}, {Goldoni},
  {G{\"o}tz}, {Mereghetti}, {Paizis}, {Segreto}, {Skinner}, {Ubertini}, \& {von
  Kienlin}}]{Malaguti03b}
{Malaguti}, G., {Bazzano}, A., {Beckmann}, V., {et~al.} 2003{\natexlab{a}},
  \aap, 411, L307

\bibitem[{{Malaguti} {et~al.}(2003{\natexlab{b}}){Malaguti}, {Bazzano}, {Bird},
  {Di Cocco}, {Foschini}, {Laurent}, {Segreto}, {Stephen}, \&
  {Ubertini}}]{Malaguti03a}
{Malaguti}, G., {Bazzano}, A., {Bird}, A.~J., {et~al.} 2003{\natexlab{b}},
  \aap, 411, L173

\bibitem[{{Malzac} {et~al.}(2008){Malzac}, {Lubi\'nski}, {Zdziarski}, {Cadolle
  Bel}, {T\:urler}, \& {Laurent}}]{Malzac08}
{Malzac}, J., {Lubi\'nski}, P., {Zdziarski}, A., {et~al.} 2008, \aap, 492, 527

\bibitem[{{Marcinkowski} {et~al.}(2006){Marcinkowski}, {Denis}, {Bulik},
  {Goldoni}, {Laurent}, \& {Rau}}]{Marcinkowski06}
{Marcinkowski}, R., {Denis}, M., {Bulik}, T., {et~al.} 2006, \aap, 452, 113

\bibitem[{{Paul} {et~al.}(1991){Paul}, {Ballet}, {Cantin}, {Cordier},
  {Goldwurm}, {Lambert}, {Mandrou}, {Chabaud}, {Ehanno}, \& {Lande}}]{Paul91}
{Paul}, J., {Ballet}, J., {Cantin}, M., {et~al.} 1991, Advances in Space
  Research, 11, 289

\bibitem[{{Roques} \& {Jourdain}(2005)}]{Roques05}
{Roques}, J.-P. \& {Jourdain}, E. 2005, {H}ow to analyse compact sources with
  SPI in complex cases, http://sigma-2.cesr.fr/spi/analysis/IUG.pdf

\bibitem[{{Rothschild} {et~al.}(1998){Rothschild}, {Blanco}, {Gruber},
  {Heindl}, {MacDonald}, {Marsden}, {Pelling}, {Wayne}, \&
  {Hink}}]{Rothschild98}
{Rothschild}, R.~E., {Blanco}, P.~R., {Gruber}, D.~E., {et~al.} 1998, \apj,
  496, 538

\bibitem[{{Ruiz} {et~al.}(1994){Ruiz}, {Porras}, {Ferrero}, {Reglero},
  {Sanchez}, {Lei}, {Bird}, {Carter}, \& {Dean}}]{Ruiz94}
{Ruiz}, J.~A., {Porras}, E., {Ferrero}, J.~L., {et~al.} 1994, \apjs, 92, 683

\bibitem[{{Segreto} {et~al.}(2003){Segreto}, {Labanti}, {Bazzano}, {Bird},
  {Celesti}, \& {Marisaldi}}]{Segreto03}
{Segreto}, A., {Labanti}, C., {Bazzano}, A., {et~al.} 2003, \aap, 411, L215

\bibitem[{{Share} \& {Murphy}(2001)}]{Share01}
{Share}, G.~H. \& {Murphy}, R.~J. 2001, \jgr, 106, 77

\bibitem[{{Skinner}(2008)}]{Skinner08}
{Skinner}, G.~K. 2008, \ao, 47, 2739

\bibitem[{SpaceWeather.com(2007)}]{Space07}
SpaceWeather.com. 2007, http://www.spaceweather.com/solarflares/topflares.html

\bibitem[{{Takahashi} {et~al.}(2007){Takahashi}, {Abe}, {Endo}, {Endo}, {Ezoe},
  {Fukazawa}, {Hamaya}, {Hirakuri}, {Hong}, {Horii}, {Inoue}, {Isobe}, {Itoh},
  {Iyomoto}, {Kamae}, {Kasama}, {Kataoka}, {Kato}, {Kawaharada}, {Kawano},
  {Kawashima}, {Kawasoe}, {Kishishita}, {Kitaguchi}, {Kobayashi}, {Kokubun},
  {Kotoku}, {Kouda}, {Kubota}, {Kuroda}, {Madejski}, {Makishima}, {Masukawa},
  {Matsumoto}, {Mitani}, {Miyawaki}, {Mizuno}, {Mori}, {Mori}, {Murashima},
  {Murakami}, {Nakazawa}, {Niko}, {Nomachi}, {Okada}, {Ohno}, {Oonuki}, {Ota},
  {Ozawa}, {Sato}, {Shinoda}, {Sugiho}, {Suzuki}, {Taguchi}, {Takahashi},
  {Takahashi}, {Takeda}, {Tamura}, {Tamura}, {Tanaka}, {Tanihata}, {Tashiro},
  {Terada}, {Tominaga}, {Uchiyama}, {Watanabe}, {Yamaoka}, {Yanagida}, \&
  {Yonetoku}}]{Takahashi07}
{Takahashi}, T., {Abe}, K., {Endo}, M., {et~al.} 2007, \pasj, 59, 35

\bibitem[{{Ubertini} {et~al.}(2003){Ubertini}, {Lebrun}, {Di Cocco}, {Bazzano},
  {Bird}, {Broenstad}, {Goldwurm}, {La Rosa}, {Labanti}, {Laurent}, {Mirabel},
  {Quadrini}, {Ramsey}, {Reglero}, {Sabau}, {Sacco}, {Staubert}, {Vigroux},
  {Weisskopf}, \& {Zdziarski}}]{Ubertini03}
{Ubertini}, P., {Lebrun}, F., {Di Cocco}, G., {et~al.} 2003, \aap, 411, L131

\bibitem[{{Vedrenne} {et~al.}(2003){Vedrenne}, {Roques}, {Sch{\"o}nfelder},
  {Mandrou}, {Lichti}, {von Kienlin}, {Cordier}, {Schanne}, {Kn{\"o}dlseder},
  {Skinner}, {Jean}, {Sanchez}, {Caraveo}, {Teegarden}, {von Ballmoos},
  {Bouchet}, {Paul}, {Matteson}, {Boggs}, {Wunderer}, {Leleux},
  {Weidenspointner}, {Durouchoux}, {Diehl}, {Strong}, {Cass{\'e}}, {Clair}, \&
  {Andr{\'e}}}]{Vedrenne03}
{Vedrenne}, G., {Roques}, J.-P., {Sch{\"o}nfelder}, V., {et~al.} 2003, \aap,
  411, L63

\bibitem[{{Weidenspointner} {et~al.}(2003){Weidenspointner}, {Kiener}, {Gros},
  {Jean}, {Teegarden}, {Wunderer}, {Reedy}, {Atti{\'e}}, {Diehl}, {Ferguson},
  {Harris}, {Kn{\"o}dlseder}, {Leleux}, {Lonjou}, {Roques}, {Sch{\"o}nfelder},
  {Shrader}, {Sturner}, {Tatischeff}, \& {Vedrenne}}]{Weidenspointner03}
{Weidenspointner}, G., {Kiener}, J., {Gros}, M., {et~al.} 2003, \aap, 411, L113

\bibitem[{{Winkler}(1994)}]{Winkler94}
{Winkler}, C. 1994, \apjs, 92, 327

\bibitem[{{Winkler} {et~al.}(2003){Winkler}, {Courvoisier}, {Di Cocco},
  {Gehrels}, {Gim{\'e}nez}, {Grebenev}, {Hermsen}, {Mas-Hesse}, {Lebrun},
  {Lund}, {Palumbo}, {Paul}, {Roques}, {Schnopper}, {Sch{\"o}nfelder},
  {Sunyaev}, {Teegarden}, {Ubertini}, {Vedrenne}, \& {Dean}}]{Winkler03}
{Winkler}, C., {Courvoisier}, T.~J.-L., {Di Cocco}, G., {et~al.} 2003, \aap,
  411, L1

\end{thebibliography}

\begin{appendix} 
\label{apa}

\section{Tests of spectral extraction methods applied to the ISGRI detector}

The main aim of the tests presented here was an examination of the behaviour of 
different spectral extraction techniques in a situation when the source emission 
is very weak.  The standard OSA software for ISGRI spectral extraction is the
PIF-based \sl ii\_spectra\_extract \rm tool, using the maximum likelihood 
technique with Poissonian distribution \citep{Goldwurm03}. A detailed description 
of this technique will be published soon (A. Goldwurm and A. Gros, in 
preparation). Besides the OSA 7.0 standard method and the PPDF method applied
to the ISGRI spectral extraction, five alternative methods were checked. Two of 
them also represent a Bayesian approach with the direct handling of probability 
density functions, where the Poisson distribution is replaced by the Gaussian PDF 
(hereafter GPDF). The first one holds the Gaussian width parameter $\sigma$ equal 
to the square root of the number of counts given by the model, whereas the second 
uses instead a $\sigma$ equalling the square root of the simulated number of 
counts. As in the case of the PPDF method, the final outcome of these two 
techniques is the probability density distribution for the source net count rate. 
Then, two standard statistics methods were controlled, where the test 
statistic was either a $\chi^{2}$ (Gaussian likelihood) or C statistic 
(Poisson likelihood). For these two techniques the result collected for a single 
science window consisted of only two numbers: the source count rate minimizing 
the $\chi^{2}$ (or C) and its uncertainty corresponding to a change of the 
statistic by a certain value. In the case of the $\chi^{2}$ statistic, both 
variants of the Gaussian width parameter (from the model or from data) were 
checked. When the number of counts simulated for a given pixel was zero, the 
`data variant' $\sigma$ was set to be 1.  

Final results for the methods producing probability density distributions were 
computed using the product of all individual PDFs obtained for single science
windows constituting a given data set (simulation run). OSA spectra were summed
with a dedicated tool $spe\_pick$ (included in the OSA software). In the case of 
methods based on $\chi^{2}$ and C statistic, the results for a single science 
window were computed over a grid of the source count rate and background 
normalization values. This was more time demanding, but allowed us to avoid 
potential problems with the convergence of the fitting procedure when a global 
minimum (maximum) is searched for. Uncertainties in a single science window 
result were determined by finding the width of the confidence region projected 
onto the source count-rate axis, for which the $\chi^{2}$ or C statistic value 
changed by 1.0, i.e. 1-sigma confidence level for one interesting parameter 
\citep{Cash79}. Then, the count rates for larger data sets (e.g. one or all 
simulation runs) were calculated as a mean weighted by these uncertainties. 
This approach is, of course, not justified for the C statistic because the 
likelihood obtained for a single science window is still not well approximated by 
the Gaussian distribution. In consequence, the standard weighted mean is 
inadequate for an asymmetric Poisson distribution characterizing `low count rate' 
data, where the position of the density maximum is usually far from the position 
of the mean. Because the maximum likelihood is often peaked at zero and the 
lower and upper limits are located at quite different distances from the maximum, 
the standard deviation error used for computing the weighted mean was set to the 
arithmetic mean of the lower and upper errors. 
 
\begin{table}
\begin{center}
\caption{Initial conditions of tests done with the simulated ISGRI data. The 
signal to noise ratio, S/N, was calculated for a 1 Ms effective exposure time.}
\label{simtest}
\begin{tabular}{cccc}
\hline\hline
Energy range & Source count rate & Bkg. count rate & S/N \\
\multicolumn{1}{c}{[keV]} & [cps] & [cps] & \\
\hline
 14.0--17.8 & 0.035 & 31.1 & 6.3 \\
 17.8--21.6 & 0.037 & 35.8 & 6.2 \\
 21.6--25.4 & 0.032 & 39.7 & 5.1 \\
 25.4--31.2 & 0.035 & 44.1 & 5.3 \\
 31.2--37.9 & 0.026 & 35.7 & 4.4 \\
 37.9--45.6 & 0.022 & 34.9 & 3.7 \\
 45.6--55.1 & 0.021 & 48.9 & 3.0 \\
 55.1--68.5 & 0.021 & 92.0 & 2.2 \\
 68.5--85.8 & 0.013 & 68.5 & 1.6 \\
 85.8--112.6 & 0.013 & 65.1 & 1.6 \\
 112.6--166.2 & 0.0083 & 85.6 & 0.9 \\
 166.2--276.3 & 0.0023 & 105.2 & 0.22 \\
 276.3--462.5 & 0.00033 & 69.9 & 0.04 \\
\hline\hline 
\end{tabular}  
\end{center}
\end{table}

Simulations were performed for the ISGRI detector with the aim of comparing the 
PPDF method with the standard OSA ISGRI results and because the method presented 
here was planned to be used with the high-energy ISGRI data. The observations of 
the Cassiopeia field made in Revs. 0384--0395 served as a simulation template. 
They included 428 science windows with about 880 ks total exposure time. A fake 
source of 1 mCrab strength was assumed to be at the Cas A position. The standard 
OSA 7.0 PIFs computed for Cas A and background maps were used to simulate the 
shadowgrams in thirteen energy bands listed in Table \ref{simtest}. The simulated 
source and background count rates are also presented, together with the 
signal-to-noise ratios (S/N) for 1 Ms exposure time. The noise was estimated 
simply to be the ratio of the square root of background counts to the exposure 
time.

Because the simulated count rates were too low to achieve a 3-sigma detection in 
the higher energy bands, simulations were repeated 20 times, resulting in 
17.6 Ms of total simulated exposure time. In addition, to check the case with
conditions more relevant to PICsIT, another set of simulations with the 
background count rates assumed to be 10 times higher than those of ISGRI was 
performed.

\begin{figure}[!h]
\centering
\includegraphics[width=9cm]{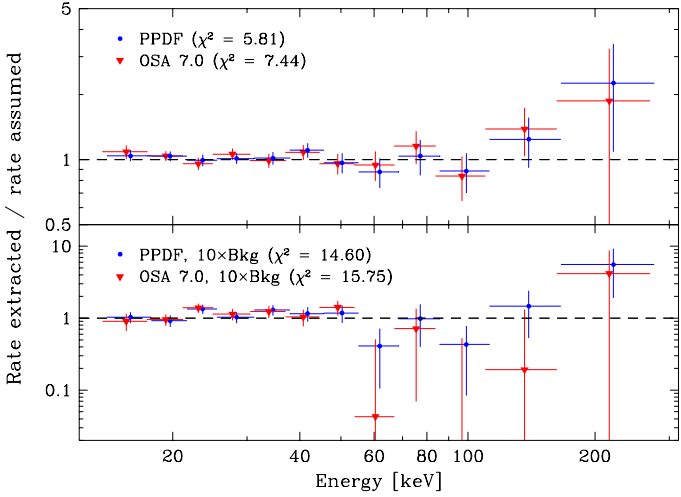}
\caption{Results of tests (all 20 simulation runs merged together) for 
the two best spectral extraction methods: the new PPDF technique and the 
standard ISGRI spectral extraction software released in OSA 7.0. Upper panel: 
ratios between the extracted and assumed count rates for the normal ISGRI 
background level. Lower panel: ratios obtained for a background 10 times higher. 
The energy bands of OSA results are shifted slightly to the left for a better 
visibility.}
\label{best}
\end{figure}

Figure \ref{best} presents the main result of the tests. For the standard 
background level, both PPDF and OSA 7.0 methods reproduce correctly the source 
count rates with at least a 3-sigma significance for energy bands below 
166 keV. The PPDF method performs more reliably, producing results very close to 
the assumed values in all bands below 40 keV. Above that energy, PPDF count rates 
are, on average, closer to the true value than the OSA 7.0 count rates. In the 
case of a test made with a background level higher by a factor of 10, the 
differences between the PPDF and OSA 7.0 results are small in the 18--55 
keV energy range, although, at higher energies the PPDF method appears to 
estimate the count rate more reliably (note that there is still insufficient data 
to obtain a detection at higher energy). 

\begin{figure}[!h]
\centering
\includegraphics[width=9cm]{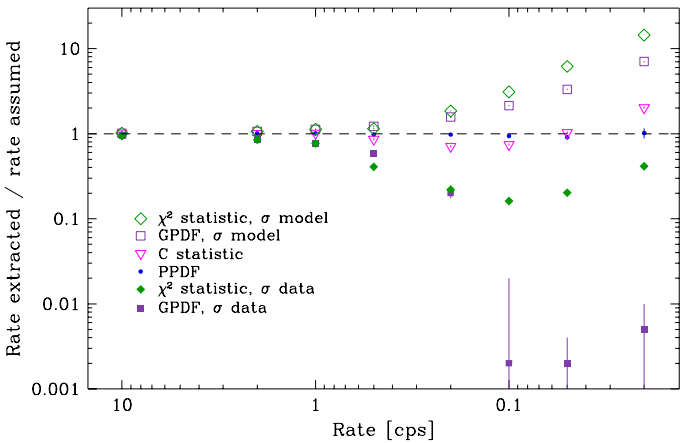}
\caption{Comparison between the results of the best method (PPDF) and the 
results obtained with the use of several alternative methods. Tests were done for 
the 17.8--21.6 keV energy band and the standard ISGRI background. One simulation
run (428 science windows) was done for the count rates above 0.2 cps. For 0.2, 
0.1, 0.05 and 0.02 cps there were 2, 5, 10 and 10 simulation runs merged 
together, respectively.}
\label{rates}
\end{figure}

Since alternative methods were not expected to produce good results at low count 
rates, their behaviour was tested first for a wider range of simulated count 
rates. Figure \ref{rates} shows the results of this test for the second energy 
band (17.8--21.6 keV) with the count rates decreasing from 10 to 0.02 cps. The 
GPDF and $\chi^{2}$ statistic methods using the model $\sigma$ parameter diverge 
from the true result by a factor of 2 or more starting from 0.2 cps, producing 
too high count rates. When the data $\sigma$ parameter is used instead in these 
two approaches, the reproduced count rates are too small by a factor of $\geq$ 2 
already at 0.5 cps. The C statistic performs much better than the methods 
based on the Gaussian distribution: deviations from the true value are 
observed below 0.5 cps but they remain small ($<$ 20\%), except for the lowest 
tested count rate (0.02 cps). Below 0.05 cps, the count rates reproduced by the 
GPDF, the $\chi^{2}$ statistic (both with the model $\sigma$) and the C statistic 
all saturate at different levels. We checked with an additional test that these 
levels are proportional to the background level: for a 10 times higher background  
they are about 3 times higher, in accordance with the expected increase in the 
standard deviation, equal to the square root of the total count number.       

\begin{figure}[!h]
\centering
\includegraphics[width=9cm]{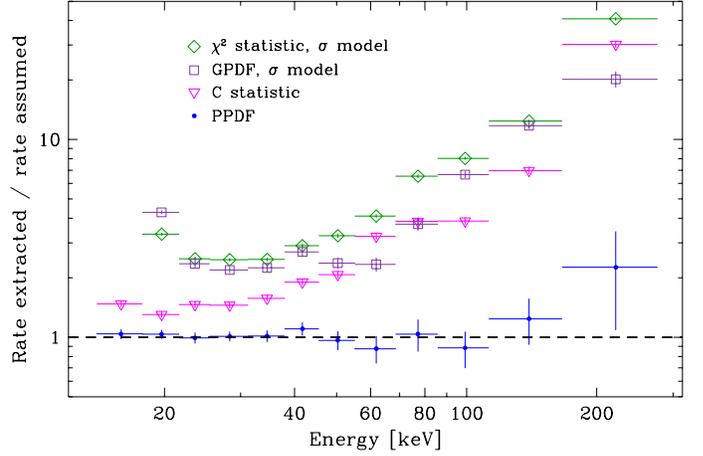}
\caption{Comparison between the results of the best method (PPDF) and the results 
obtained with the use of several alternative methods. Tests were done for the
standard ISGRI background.}
\label{methods}
\end{figure}

Figure \ref{methods} presents again the comparison between the PPDF and other
methods, this time for all tested energy bands and for the count rates 
corresponding to the 1 mCrab source observed with ISGRI. At this count rate
level, only the PPDF technique provides the correct results. The deviations
observed for the other methods are correlated with the increasing background 
level (see Table \ref{simtest}). At low energy (below 20 keV), when the number 
of counts simulated for a given pixel is often equal to zero (due to the low 
energy thresholds set for the pixels), both methods using the Gaussian 
approximation (GPDF and $\chi^{2}$) fail completely.   

\begin{figure}[!h]
\centering
\includegraphics[width=9cm]{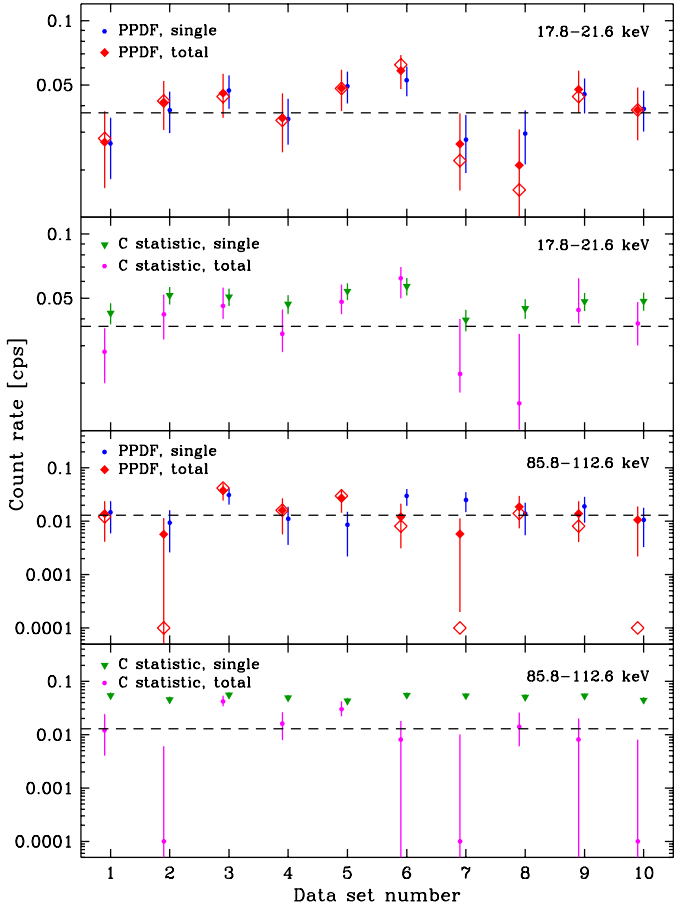}
\caption{Results of two variants of the C statistic method compared to the 
corresponding results for the PPDF method. The comparison is made for two energy 
bands representing high (at low energy) and low (at high energy) S/N ratios. In 
each panel the points shifted to the right illustrate the results of a given 
method applied to each science window separately and then merged by multiplying 
all PDFs (PPDF) or by computing the weighted mean (C statistic). Points shifted 
to the left represent the variant of applying either method to the total data set 
at once. Open diamonds indicate the position of the maximum of the PPDF 
distribution. Dashed lines show the count rate assumed in simulations.}
\label{multi}
\end{figure}

Figure \ref{multi} compares the results of the PPDF and C statistic methods 
applied to single science windows and merged later in the way described above, 
and the results obtained when all the data from the entire simulation run (428 
science windows) were extracted at once with these methods. Merging the data
is possible because simulations are always performed with the same 
background level assumed and all 428 data sets can be treated as one set, 
corresponding to the detector with 428 times more pixels. Contrary to the 
standard procedure, the background normalization was found once for all science 
windows. For both higher and lower S/N bands (low and high energy, respectively), 
both PPDF approaches give similar results, completely consistent, on average, 
with the assumed count rate. Direct application of the PPDF count rate extraction 
to the data of all science windows at once leads to slightly larger final errors, 
because of a broader two-dimensional PDF associated with a broader background 
count rate distribution in a larger simulation set. It seems more reliable to 
integrate the two-dimensional PDF over the background normalization factor for 
each science window separately because it allows a more careful control of 
background fluctuations. Since in a real situation the instrumental background 
varies quickly, this strategy should be followed and one has to work on possibly 
short-time data sets before merging the results from a longer observation period.

On the other hand, the C statistic technique produces the correct results 
only when applied to a larger data set. In this case, the maximum likelihood 
method of course finds a correct position of the PDF maximum, as shown in Fig. 
\ref{multi}, where the PPDF maxima are marked for comparison. Nevertheless, it 
appears to be rather inefficient for low S/N data, as shown in the lowest panel 
of Fig. \ref{multi}. This is a consequence of using the maximum, which is quite  
often located far from the gravity centre of the asymmetric Poisson PDF. In 
principle, when the count rates are not too small, the C statistic applied to 
all data at once can be used, provided that there is no need to update the 
result when a new data set arrives. For such an update, the PDF methods storing 
all distributions derived for single science windows are faster because in this 
case one has only to extract the PDFs for the new data and then merge them with 
the older results. Also any change in the data selection, e.g. for the source 
variability studies, makes the method based on C statistic rather impractical.

Besides the performance tests described already, the two most reliable methods 
were in addition checked with respect to the scatter in the results and 
the correctness of the derived uncertainties. These tests were done using the 
results of 20 simulation runs and merging them in a random way to obtain a new 
simulation run, again with 428 science windows. For each new run, each science 
window was selected randomly from one of 20 runs, to preserve the original 
science-window content of the simulation set. This procedure was repeated 500 
times, to calculate $\chi^{2}$, arithmetic mean of the standard deviations 
$\bar{\sigma}$, and the fraction of results containing the true count-rate value 
within the uncertainty limits. The complete results of these tests are listed in 
Table \ref{coverage},  for the first 11 energy bins and for three energy bins 
(4, 7, and 10), for the default and 10 times higher background, respectively. 

Both methods were examined with the so-called `frequentist coverage' 
test, where the fraction of results covering the true count rate value by error 
bars was computed. In case of the PPDF method, there are many ways to determine 
the uncertainty, so-called `credible intervals' in Bayesian nomenclature, because 
the PDF associated with the result is asymmetric. Here, three variants were 
tested. The first one was a standard deviation defined as the square root of the 
variance determined by integrating $(s-\bar{s})^{2}$ (where $s$ is the source 
count rate) weighted by the PDF, i.e. the second central moment of the PDF. The 
second variant was the shortest interval containing the desired fraction 
of the PDF integral, 68.3\% for 1-$\sigma$ error. The third option was the 
central credible interval, found as limits above and below which the PDF 
integrals were equal to (1-0.683)/2. 

The standard deviation and central credible interval approaches gave almost
identical results and thus only the former is shown in Table \ref{coverage}. They 
tend to overestimate slightly the uncertainty, whereas the shortest interval 
approach seems to underestimate the uncertainty. Nevertheless, all of these 
variants produce errors that are, in general, in agreement with the 0.683 
coverage probability taking into account about 5\% deviation expected for 500 
simulations performed. The OSA 7.0 uncertainties appear to pass the coverage test 
more successfully than those obtained with the PPDF technique. However, this 
happens at the cost of much larger error values (measured by $\bar{\sigma}$) and 
a larger scatter in the results. Plainly the PPDF method is more precise than the 
method implemented in the OSA software. 

\begin{table}
\begin{center}
\setlength{\tabcolsep}{1.8mm}
\caption{Uncertainty tests done for the OSA 7.0 and PPDF spectral extraction 
methods. $\chi ^{2}$ is summed over all 500 shots, $\bar{\sigma}$ is an 
arithmetic mean of the standard deviation errors, $f_{1\sigma}$ is a fraction 
of shots covering the true count rate value inside the error limits and $f_{s}$ 
is a similar fraction of shots computed for the shortest credible interval errors
for the PPDF method. Three rows at the bottom present the results of the same 
test made for the background assumed to be 10 times higher than the default one.}
\label{coverage}
\begin{tabular}{cccccccc}
\hline\hline
Energy range & \multicolumn{3}{c}{OSA 7.0} & \multicolumn{4}{c}{PPDF} \\
\multicolumn{1}{c}{[keV]} & $\chi^{2}$ & $\bar{\sigma}$ & $f_{1 \sigma}$ & 
$\chi^{2}$ & $\bar{\sigma}$ & $f_{1 \sigma}$ & $f_{s}$ \\
\hline
 14.0--17.8   & 471 & 0.0114 & 0.69 & 425 & 0.0089 & 0.73 & 0.72 \\
 17.8--21.6   & 566 & 0.0087 & 0.65 & 537 & 0.0084 & 0.66 & 0.64 \\
 21.6--25.4   & 467 & 0.0086 & 0.67 & 443 & 0.0084 & 0.71 & 0.71 \\
 25.4--31.2   & 478 & 0.0090 & 0.68 & 503 & 0.0087 & 0.68 & 0.68 \\
 31.2--37.9   & 481 & 0.0081 & 0.72 & 459 & 0.0078 & 0.73 & 0.71 \\
 37.9--45.6   & 513 & 0.0080 & 0.67 & 569 & 0.0076 & 0.65 & 0.63 \\
 45.6--55.1   & 489 & 0.0094 & 0.70 & 486 & 0.0085 & 0.72 & 0.68 \\
 55.1--68.5   & 508 & 0.0130 & 0.69 & 443 & 0.0105 & 0.74 & 0.56 \\
 68.5--85.8   & 447 & 0.0112 & 0.71 & 344 & 0.0088 & 0.78 & 0.65 \\
 85.8--112.6  & 462 & 0.0109 & 0.71 & 285 & 0.0083 & 0.82 & 0.61 \\
 112.6--166.2 & 460 & 0.0125 & 0.70 & 459 & 0.0091 & 0.71 & 0.75 \\
\hline
 25.4--31.2   & 537 & 0.0284 & 0.71 & 348 & 0.0225 & 0.78 & 0.66 \\
 45.6--55.1   & 487 & 0.0299 & 0.67 & 435 & 0.0221 & 0.72 & 0.75 \\
 85.8--112.6  & 513 & 0.0344 & 0.65 & 367 & 0.0207 & 0.76 & 0.88 \\
\hline\hline 
\end{tabular}  
\end{center}
\end{table}

\begin{figure}[!h]
\centering
\includegraphics[width=9cm]{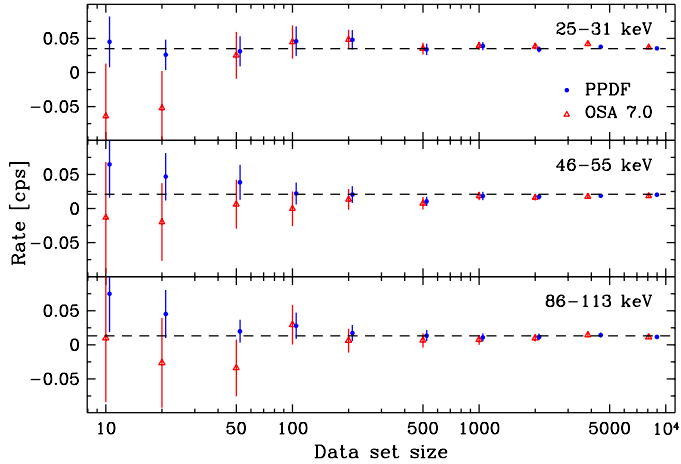}
\caption{Results of tests for the two best spectral extraction methods, PPDF and 
OSA 7.0. Progress in the accuracy of the extracted count rates with the 
increasing  number of simulated science windows  is presented for several 
energy bands.}
\label{progress}
\end{figure}

Yet another comparison between the results of OSA 7.0 and the PPDF methods is 
shown in Fig. \ref{progress}. Progress in the determination of the source count 
rate is presented for three energy bands. The PPDF produces overestimated count 
rates when there is insufficient information in the data, as a consequence of 
broad but non-negative PDFs. On the other hand, the OSA 7.0 software has some 
tendency to underestimate count rates when the data set remains too small for a 
detection. With the increasing amount of data both approaches exhibit normal 
random oscillations around the true value, with the amplitude of variations 
decreasing with the data set size. As already demonstrated in Table 
\ref{coverage}, the scale of oscillations is clearly smaller in the case of the 
PPDF method, converging quite rapidly to the true value.

Extensive tests performed with simulated data showed that only two count-rate
extraction methods work correctly for data of very low signal-to-noise ratio. 
The method handling the Poisson probability functions can be treated as a 
reference, because it allows us to obtain the most precise and quickly converging 
results. The only limitation of this method is a long computation time  
when many parameters have to be treated at once. However, in the case of high 
energy astrophysics the number of sources in the field of view is always small 
and this limitation poses no serious problem. The technique implemented in the 
OSA software also produces very good results and can be recommended for general 
use whenever computation speed and robustness is needed. Among the other methods 
tested, only that based on the C statistic performs relatively well when the 
signal is not extremely weak. Because it can produce systematically 
underestimated results, its performance in a given application should be tested 
with simulated data.

\end{appendix}

\end{document}